\documentclass[pra,aps,superscriptaddress,showpacs,floatfix,tightenlines,twocolumn]{revtex4-1}
\usepackage{graphicx}
\usepackage{dcolumn}
\usepackage{bm}
\usepackage{amssymb}
\usepackage{amsmath}
\usepackage{url}
\usepackage{amsthm}
\usepackage{layout}
\usepackage{epsfig}
\usepackage{graphicx}
\usepackage{epstopdf}
\usepackage{booktabs}
\usepackage{float}
\usepackage{subfigure}
\usepackage[hyperindex,breaklinks]{hyperref}
\usepackage{xcolor}

\begin{document}

\title{Gaussian unsteerable  channels and computable quantifications of Gaussian  steering}
\author{Taotao Yan}
\affiliation{College of Mathematics, Taiyuan University of Technology, Taiyuan, 030024, China}
\author{Jie Guo}
\affiliation {College of Mathematics, Taiyuan University of Technology, Taiyuan, 030024, China}
\author{Jinchuan Hou}
\email{
houjinchuan@tyut.edu.cn, jinchuanhou@aliyun.com}
\affiliation{College of Mathematics, Taiyuan University of Technology, Taiyuan, 030024, China}
\author{Xiaofei Qi}
\email{
qixf1981@126.com}
\affiliation{School of Mathematics and Statistics, Shanxi University,
Taiyuan 030006, China}
\author{Kan He}
\email{hekanquantum@163.com}
\affiliation{College of Mathematics, Taiyuan University of Technology, Taiyuan, 030024, China}

\date{\today }
\begin{abstract}
The current quantum resource theory for Gaussian steering  in continuous-variable systems is flawed and incomplete. Its primary shortcoming stems from an inadequate comprehension of the architecture of Gaussian channels transforming Gaussian unsteerable  states into Gaussian unsteerable  states, resulting in a restricted selection of free operations.
In the present paper, we explore in depth the structure of such $(m+n)$-mode  Gaussian channels,  and introduce the class of the Gaussian unsteerable channels and the class of maximal Gaussian unsteerable channels, both of them may be chosen as the
 free operations, which completes the resource theory for Gaussian steering from $A$ to $B$  by Alice's Gaussian measurements. We also propose two  quantifications $\mathcal{J}_{j}$ $(j=1,2)$ of $(m+n)$-mode Gaussian steering from $A$ to $B$. The computation of the value of $\mathcal{J}_{j}$ is straightforward and efficient, as it solely relies on the covariance matrices of  Gaussian states, eliminating the need for any  optimization procedures. Though $\mathcal{J}_{j}$s are not genuine Gaussian steering measures, they have some nice properties such as   non-increasing under certain Gaussian unsteerable  channels. Additionally,  we compare ${\mathcal J}_2$ with the Gaussian steering measure $\mathcal N_3$, which is based on the Uhlmann fidelity, revealing that ${\mathcal J}_2$ is an upper bound of $\mathcal N_3$ at certain class of  $(1+1)$-mode Gaussian pure states. As an illustration, we apply $\mathcal J_2$ to discuss the behaviour of Gaussian steering for a special class of $(1+1)$-mode Gaussian states in Markovian environments, which uncovers the intriguing phenomenon of rapid decay in quantum steering.

\end{abstract}

\maketitle


\section{Introduction}

The Einstein-Podolsky-Rosen steering is a nonlocal property of quantum states which is stronger than the entanglement \cite{EPR} and weaker than the Bell nonlocality \cite{WJD}, and has attracted a great deal of interest during the past decades. In order for quantum steering to be useful, the first task is to be able to detect it in experiments, and to our knowledge, the detection of steering has witnessed remarkable advancements. The second task is to measure the steering for practical purposes. In finite-dimensional systems, the measures of steering have garnered extensive research attention \cite{MJ}-\cite{THX}, leading to increased understanding of their potential usefulness for tasks that harness steering as a fundamental resource  \cite{GA}.

The continuous-variable (CV) quantum systems are also fundamental important from theoretical and experimental views. {In particular,
 Gaussian states can be readily produced and manipulated in experimental settings.}  One can therefore consider the special class of Gaussian quantum resource
theories, whose free states and operations are required to be Gaussian.  For instance, taking  Gaussian separable states as free states and Gaussian local operations and classic communications (GLOCCs) as free operations makes Gaussian entanglement into a quantum resource \cite{LRX}. Such an effort has also been performed in \cite{LRX} for Gaussian steering where it was claimed that the Gaussian steering is a quantum resource. Unfortunately, there are two points of ambiguity in \cite{LRX}. The first one is that, in the discussion of the standard assumption of free states and in the context of bipartite Gaussian steering, the free state was defined as the intersection of the set of states that are unsteerable from $A$ to $B$ and the set of Gaussian states. However, in the application part of \cite{LRX}, Gaussian unsteerable state based on Gaussian measurements { was} taken as the free state by default, and the steering criterion from \cite{WJD} was applied. Note that the notion of Gaussian steerable states was initially introduced in \cite{WJD} through the Gaussian positive operator-valued measurements (GPOVMs), and a definitive and exhaustive criterion, formulated in terms of covariance matrices, was established to determine when a Gaussian state is unsteerable. It is important to realize that the Gaussian unsteerable states from $A$ to $B$ by Alice's Gaussian measurements may not necessarily be the unsteerable states from $A$ to $B$. The second
one is that, in \cite{LRX} the free operations were chosen to be the one-way GLOCCs, which constitutes only a very small subset of the Gaussian channels transforming Gaussian { unsteerable} states from $A$ to $B$ under Alice's Gaussian measurements to the same kind states. Therefore, it is crucial to revisit the question of whether Gaussian steering can be used as a quantum resource.

In the present paper, we  complete
the resource theory for Gaussian steering from $A$ to $B$ by Alice's Gaussian measurements. The free states are Gaussian unsteerable states. There are two  reasonable choices of the free operations:  {\it the Gaussian unsteerable  channels} or {\it the maximal Gaussian unsteerable channels}.  A maximal  Gaussian unsteerable   channel is a Gaussian channel which transforms Gaussian unsteerable states into Gaussian unsteerable states. A Gaussian unsteerable  channel is a special kind of maximal Gaussian  unsteerable channel which is easily recognized, constructed and applied  in quantum information process.
  The set of all Gaussian unsteerable   channels contains all two-way  GLOCCs and is a proper (but large) subset of the set of all maximal Gaussian unsteerable   channels. Several Gaussian steering measures based on the distances between two Gaussian states are proposed, which ensure that the Gaussian steering is a quantum resource. However, these measures are { difficult} to be calculated.

In order to apply Gaussian quantum steering in the quantum information scenario, it is useful  to propose  quantifications of Gaussian steering that are easily calculated  for all $(m+n)$-mode bipartite Gaussian states, though they may not be true Gaussian steering measures. In recent years, some efforts have been devoted to  quantifying the quantum steering for bipartite Gaussian states. In \cite{KA}, Kogias and Adesso proposed a quantification of the Gaussian
steering of $(1+1)$-mode Gaussian states based on the EPR paradox. However, this quantification is applicable only to $(1+1)$-mode Gaussian states.   Kogias, Lee, Ragy, et al. in \cite{KLRA} proposed another quantification of the Gaussian steering for any bipartite Gaussian state based on the symplectic eigenvalues of the matrix.  So far, beyond the two quantifications mentioned above, surprisingly, there have been no other quantifications for Gaussian steering.

 The second aim of the present paper is to propose two accessible approaches to quantifying steering for bipartite Gaussian states  (by Gaussian measurements on Alice's side). These quantifications of Gaussian steering relate only to the covariance matrices of Gaussian states and avoid the conventional optimization process, which reduces the computational complexity greatly.

This paper is organized as follows. In Section \ref{sec:2}, we review the concepts of Gaussian states, Gaussian quantum steering, and a known quantum steering criterion for Gaussian states. In Section \ref{sec:3}, we present an operational framework for Gaussian steering from $A$ to $B$ by Alice's Gaussian measurements as a quantum resource. In Section \ref{sec:4}, we define two quantifications $\mathcal{J}_{j}$ $(j=1,2)$ of the Gaussian steering for any $(m+n)$-mode Gaussian states, and present some properties of these quantifications. In addition, we give the expressions of these
quantifications for $(m+n)$-mode Gaussian pure state with a standard form of the covariance matrix and a special class of $(1+1)$-mode Gaussian states. In Section \ref{sec:5}, for $(1+1)$-mode Gaussian pure state, we  compare $\mathcal{J}_{2}$ with the Gaussian steering measure $\mathcal{N}_{3}$ based on the Uhlmann fidelity.
In Section \ref{sec:6}, we discuss the behaviour of the Gaussian steering by using the quantification ${\mathcal J}_2$ for a special class of $(1+1)$-mode Gaussian states in Markovian environments. Section \ref{sec:7} is a short conclusion. All mathematical proofs are presented in Appendix section.

\section{Preliminaries}\label{sec:2}
In this section, we briefly review some notions and notations concerning Gaussian states
and Gaussian quantum steering as well as  a criterion for Gaussian
steering.
\subsection{Gaussian states}
For an arbitrary state $\rho$
in an $n$-mode CV system with state space $H$, $\rho$
is called a Gaussian state  if its characteristic function
$\chi_{\rho}(z)$ has the form
$$\chi_{\rho}(z)={\rm Tr}(\rho W(z))=\exp[-\frac{1}{4}z^{\rm T}\Gamma z+i{\mathbf
d}^{\rm T}z],\eqno(2.1)
$$
where $z=(x_{1},y_{1},\cdots,x_{n},y_{n})^{\rm T}\in\mathbb{R}^{2n}$, $W(z)={\rm exp}(iR^{\rm T}z)$ is Weyl operator,
 $$\begin{array}{rl} {\mathbf d}=&(\langle\hat R_1 \rangle, \langle\hat R_2
\rangle, \ldots ,\langle\hat R_{2n} \rangle)^{\rm T}\\=&({\rm Tr}(\rho
\hat{R}_1), {\rm Tr}(\rho \hat{R}_2), \ldots, {\rm Tr}(\rho \hat{R}_{2n}))^{\rm
T}\in{\mathbb R}^{2n}\end{array}$$ is  the mean vector of $\rho$ and
$\Gamma=(\gamma_{kl})\in \mathcal{M}_{2n}(\mathbb R)$ is the covariance matrix of $\rho$ defined by $$\gamma_{kl}={\rm Tr}[\rho
(\Delta\hat{R}_k\Delta\hat{R}_l+\Delta\hat{R}_l\Delta\hat{R}_k)],$$
in which, $\Delta\hat{R}_k=\hat{R}_k-\langle\hat{R}_k\rangle$,
$\langle\hat{R}_k\rangle={\rm Tr}[\rho\hat{R}_k]$,
$R=(\hat{R}_1,\hat{R}_2,\cdots,\hat{R}_{2n})=(\hat{Q}_1,\hat{P}_1,\cdots,\hat{Q}_n,\hat{P}_n)$.
$\hat{Q_k}=(\hat{a}_k+\hat{a}_k^\dag)/\sqrt{2}$,
$\hat{P_k}=-i(\hat{a}_k-\hat{a}_k^\dag)/\sqrt{2}$ ($k=1,2,\cdots,n$)
are  respectively the position and momentum operators in the $k$th mode,
$\hat{a}_k^\dag$ and $\hat{a}_k$ are respectively the creation and annihilation
operators in the $k$th mode \cite{SP,GJ}. Here, as usal, $\mathcal{M}_d(\mathbb R)$
stands for the algebra of all $d\times d$ matrices over the real
field $\mathbb R$. Note that  $\Gamma$ is real symmetric and
satisfies the condition $\Gamma +i\Omega_{n}\geq 0$, where
$\Omega_n= \underbrace{\Omega\oplus \Omega\cdots\oplus\Omega}_n\in{\mathcal M}_{2n}(\mathbb R)$ with $\Omega=\left(\begin{array}{cc} 0 & 1\\ -1 & 0 \end{array}\right)$. By (2.1), every Gaussian state $\rho$ is determined by its covariance matrix $\Gamma$ and mean vector ${\mathbf d}$, and thus, one can write $\rho=\rho(\Gamma, {\mathbf d})$.

 The  covariance matrix $\Gamma$  of every $(m+n)$-mode Gaussian state $\rho=\rho_{AB}$
can be written as
$$
\Gamma=\left(\begin{array}{cc}A& C\\
C^{\rm T} & B\end{array}\right),\eqno(2.2)
$$
where $C\in \mathcal{M}_{2m\times 2n}({\mathbb R})$ and $A \in \mathcal{M}_{2m}({\mathbb R})$ (resp. $B\in
\mathcal{M}_{2n}({\mathbb R})$) is the covariance matrix of the reduced state $\rho_A={\rm Tr}_B(\rho_{AB})$ (resp.
$\rho_B={\rm Tr}_A(\rho_{AB})$).
Particularly, for any $(1+1)$-mode Gaussian state $\rho$, by means of local Gaussian
unitary (symplectic at the  covariance matrix level) operations, its covariance matrix $\Gamma$ has a standard form
$$\Gamma_0=\left(\begin{array}{cc}A& C\\
C^{\rm T} & B\end{array}\right)=\left(\begin{array}{cccc}a&0&c&0\\0&a&0&d\\c&0&b&0\\
0&d&0&b\end{array}\right),\eqno(2.3)$$
where $a, b, c, d\in{\mathbb R}$, $a,b\geq1$, $a(ab-c^{2})-b\geq0$, $b(ab-d^{2})-a\geq0$, $(ab-c^{2})(ab-d^{2})+1-a^{2}-b^{2}-2cd\geq0$.

\subsection{Gaussian quantum steering}

Before reviewing the definition of the Gaussian quantum steering, let us first recall the definition of GPOVM.

\if false {\it Gaussian positive operator-valued measurement.}\fi
An
$n$-mode GPOVM
$\Pi=\{\Pi_{\alpha}\}_\alpha$  is defined  as
$$\Pi_{\alpha}=\frac{1}{\pi^n}D(\alpha)\varpi
D^\dag(\alpha),$$ where $D(\alpha)={\rm
exp}[\displaystyle\sum_{j=1}^{n}(\alpha_j\hat{a}_j^\dag-\alpha_j^*\hat{a}_j)]$ is
the $n$-mode displacement operator, $\alpha \in{\mathbb C^n}$,
$\varpi$  is  the seed state of the GPOVM $\Pi$ which is an $n$-mode Gaussian state with  zero mean vector and covariance matrix $\Sigma$. So we can denote a
GPOVM with the seed covariance matrix $\Sigma$ by
$\Pi^\Sigma=\{\Pi_{\alpha}^\Sigma\}$ \cite{GJ}. A measurement assemblage  $
\{\Pi^{\Sigma_x}_{\alpha|x}\}_{\alpha,x}$ is a collection in which {$\{\Pi^{\Sigma_x}_{\alpha|x}\}_\alpha$} is a GPOVM for each index $x$.

\if false {\it Gaussian quantum steering.}\fi
In a bipartite Gaussian steering scenario, assume that Alice and Bob share an $(m+n)$-mode Gaussian state $\rho=\rho_{AB}$ of $(m+n)$-mode bipartite CV system $H_A\otimes H_B$, where Alice holds an $m$-mode Gaussian state $\rho_A={\rm Tr}_B(\rho)$ and Bob holds an $n$-mode Gaussian state $\rho_B={\rm Tr}_A(\rho)$.  If Alice performs a set of GPOVMs
$\{\Pi^{\Sigma_x}_{\alpha|x}\}_{\alpha,x}$, then the collection of sub-normalized ``conditional Gaussian states" of Bob forms a Gaussian assemblage $\{\sigma_{\alpha|x}\}_{\alpha,x}$ with
$$\sigma_{\alpha|x}= {\rm Tr}_A [(\Pi^{\Sigma_x}_{\alpha|x}\otimes I_B){\rho}].$$
It is clear that $\int \sigma_{\alpha|x}{\rm d}\alpha =\rho_B $ for each $x$. The Gaussian state $\rho$ is said to be Gaussian unsteerable from  $A$ to $B$,  if  for any set of  GPOVMs $\{\Pi^{\Sigma_x}_{\alpha|x}\}_{\alpha,x}$, the associated  assemblage $\{\sigma_{\alpha|x}\}_{\alpha,x}$ on Bob's side
 can be explained by a Gaussian local hidden state
(GLHS) model  as follows:
$$\sigma_{\alpha|x}= \int p_\lambda p(\alpha|x,\lambda)\sigma_\lambda d\lambda,$$
where $\lambda$ is a hidden variable,
$p_\lambda$ is a Gaussian distribution in $\lambda$, $p(\alpha|x,\lambda)$ is local {\it response function}
of Alice and it is a Gaussian distribution in $\lambda$ with a mean vector equal to $\alpha$. $\{\sigma_\lambda\}$ are {\it Gaussian hidden states} of Bob and all $\sigma_\lambda$ have the same covariance matrix but have different mean vectors $\lambda$. Otherwise, if there
exist some set of GPOVMs $\{\Pi^{\Sigma_x}_{\alpha|x}\}_{\alpha,x}$ such that the associated  assemblage $\{\sigma_{\alpha|x}\}_{\alpha,x}$ do not admit such a
GLHS model, then the state $\rho$ is called Gaussian steerable
from  $A$ to $B$ by Alice's Gaussian measurements  \cite{WJD}. { From this definition}, it is clear that a Gaussian steerable state from $A$ to $B$ is steerable regarded as a bipartite state of the discrete system $H_A\otimes H_B$ and  the set of Gaussian unsteerable states is not the intersection of the set of all Gaussian states and the set of all unsteerable states.

 For the convenience, denote by ${\mathcal US}_{A\rightarrow B}^{G}={\mathcal US}_{A\rightarrow B}^{G}(m,n)$ the set of all $(m+n)$-mode Gaussian states which is Gaussian unsteerable from $A$ to $B$ by Alice's Gaussian measurements. It is proved in \cite{WJD} that $\rho\in {\mathcal US}_{A\rightarrow B}^{G}$ if and only if
$$\Gamma_{\rho}+0_{A}\oplus i\Omega_B\geq0,\eqno(2.4)$$
where $\Gamma_{\rho}$ is the covariance matrix of $\rho$,
$0_A\in{\mathcal M}_{2m}(\mathbb R)$ is the zero matrix and $\Omega_B= \Omega_n\in{\mathcal M}_{2n}(\mathbb R)$.

Similarly, one can define and discuss the
Gaussian steering from $B$ to $ A$ by Bob's Gaussian measurements. In the present paper, we focus our attention on Gaussian steering from $A$ to $ B$ by Alice's Gaussian measurements.

\section{Resource theory for Gaussian steering}\label{sec:3}

In this section, we discuss resource theory for Gaussian steering from $A$ to $B$ by Alice's Gaussian measurements.

\subsection{Gaussian unsteerable states from A to B as free states} The standard assumptions about the set of Gaussian free states were proposed in \cite{LRX}. Namely, \\
(1) The set of Gaussian free states is invariant under displacement operations.\\
(2) The set of Gaussian free states is closed under tensor products of subsystems.\\
(3) The set of Gaussian free states is closed under partial traces of subsystems.\\
(4) The set of Gaussian free states is closed under permutations of subsystems.\\
(5) The set of Gaussian free states is closed.\\
(6) The set of covariance matrices corresponding to the Gaussian free states ensemble is upward closed, that is, if $V$ is covariance matrix of the Gaussian free state and $W\geq V$, then $W$ is also a covariance matrix of some Gaussian free state.

The set ${\mathcal US}_{A\rightarrow B}^{G}$ of Gaussian unsteerable states from $A$ to $B$ by Alice's Gaussian measurements, which {\color{red} is} completely described by the covariance matrix as in Eq.(2.4), satisfies assumption (1). It is clear that if $\rho_i\in{\mathcal{US}}^G_{A_i\to B_i}$, $i=1,2,\ldots , k$, then $\rho=\rho_1\otimes \rho_2\otimes\cdots\otimes \rho_k\in {\mathcal{US}}^G_{A\to B}$, where $H_A=H_{A_1}\otimes H_{A_2}\otimes\cdots\otimes H_{A_k}$ and $H_B=H_{B_1}\otimes H_{B_2}\otimes\cdots\otimes H_{B_k}$. So, (2) is satisfied.
As the unsteering from $A$ to $B$ is a bipartite quantum correlation and is not symmetrical about the subsystems, (3)-(4) are not applied to this situation and thus are satisfied emptily.
For assumptions (5) and (6), we have the following conclusions.

{\bf Theorem 3.1.} {\it The set ${\mathcal US}_{A\rightarrow B}^{G}$ is closed.}

{\bf Theorem 3.2.} {\it The set of covariance matrices of the Gaussian unsteerable states from $A$ to $B$ by Alice's Gaussian measurements is upward closed, that is, if $V$ is covariance matrix of a Gaussian unsteerable state from $A$ to $B$ and $W\geq V$, then $W$ is also the covariance matrix of some Gaussian unsteerable state from A to B. }

In summary, the set ${\mathcal US}_{A\rightarrow B}^{G}$ of Gaussian unsteerable states from $A$ to $B$ by Alice's Gaussian measurements satisfies the rules (1)-(6).  Hence, Gaussian unsteerable states from $A$ to $B$ are eligible as  free states.

\subsection{Gaussian unsteerable channels as free operations}
By the resource theory, a free operation can be any quantum channel which always sends free states into free states. In the situation of CV systems, the free operations are those Gaussian channels which send Gaussian free states into Gaussian free states. Thus, for Gaussian steering correlation, the free operations should be Gaussian channels that send Gaussian unsteerable states from $A$ to $B$ into Gaussian unsteerable states from $A$ to $B$.

We first check when a Gaussian channel sends Gaussian unsteerable states from $A$ to $B$ into Gaussian unsteerable states from $A$ to $B$.

Recall that a Gaussian quantum channel is a quantum channel   transforming  Gaussian states into Gaussian states and can be represented by a Gaussian unitary dilation. More precisely, an $n$-mode Gaussian channel $\Phi$ can be described by $\Phi=\Phi(K,M,\bar{d})$ as follows: for any $n$-mode Gaussian state $\rho=\rho(\Gamma,{\mathbf d})$, $\Phi(\rho(\Gamma,{\mathbf d}))=\rho(\Gamma',{\mathbf d}')$ with
$${\mathbf d}'=K{\mathbf d}+\bar{d},\Gamma'=K\Gamma K^{\rm T}+M,\eqno(3.1)$$
where $\bar{d}\in\mathbb R^{2n}$, $M, K\in {\mathcal M}_{2n}(\mathbb R)$, $M=M^{\rm T}$ satisfying $M+i\Omega_{n}-iK\Omega_{n} K^{\rm T}\geq0$ \cite{WP}. Note that, we must have $M\geq 0$. In fact, as $\Omega^{\rm T}=-\Omega$, $M+i\Omega_{n}-iK\Omega_{n} K^{\rm T}\geq0$ implies $M-i\Omega_{n}+iK\Omega_{n} K^{\rm T}\geq0$, which gives $2M\geq 0$.

\if false  $\Omega_{n}=\oplus_{j=1}^n \Omega_j\in{\mathcal M}_{2n}(\mathbb R)$. with $\Omega_j=\Omega=\left(\begin{array}{cc} 0 & 1\\ -1 & 0 \end{array}\right)$.\fi

The following observation reveals that every local Gaussian channel is unsteerable ones.

{\bf Theorem 3.3.} {\it Let $H_A\otimes H_B$ describe an $(m+n)$-mode CV system. Then, for any Gaussian channels $\Phi_{A}(K_{A},M_{A},\bar{d}_{A})$ and $\Phi_{B}(K_{B},M_{B},\bar{d}_{B})$ acting on the subsystems $H_{A}$ and $H_{B}$, respectively,   $\rho\in{\mathcal US}_{A\rightarrow B}^{G}$ implies that $(\Phi_{A}\otimes\Phi_{B})\rho\in{\mathcal US}_{A\rightarrow B}^{G}$.}

Theorem 3.3 also means that considering solely one-way GLOCCs, represented by $\Phi_A\otimes I_B$, as free operations falls significantly short of being sufficient.

 Denote by $\mathcal{B}(H)$ the von Neumann algebras of all bounded linear operators on a Hilbert space $H$. Recall that a unitary operator $U\in{\mathcal B}(H)$ is Gaussian if the unitary channel induced by $U$ is Gaussian. By Theorem 3.3, the following corollary is immediate.

{\bf Corollary 3.4.} {\it   $(U_{A}\otimes U_{B}){\mathcal US}_{A\rightarrow B}^{G}(U_{A}\otimes U_{B})^{\dagger}\subseteq{\mathcal US}_{A\rightarrow B}^{G}$ holds for all Gaussian unitary operators $U_{A}\in\mathcal{B}(H_{A})$ and $U_{B}\in\mathcal{B}(H_{B})$.}

\if false
{\bf Proof.} Let $\sigma=(U_{A}\otimes U_{B})\rho (U_{A}\otimes U_{B})^{\dagger}$ and denote by $\Gamma_\rho$, $\Gamma_\sigma$ the covariance matrices of $\rho$, $\sigma$, respectively. According to the Ref.\cite{GJ},
 $\Gamma_{\sigma}=(S_{A}\oplus S_{B})\Gamma_{\rho}(S_{A}\oplus S_{B})^{\dag}$, where $S_{A}$ and $S_{B}$ are the symplectic matrices  corresponding to $U_{A}$ and $U_{B}$, respectively. In Theorem 3.3, let $M_{A}=M_{B}=0$, $K_{A}=S_{A}$ and $K_{B}=S_{B}$, then $\Gamma_{\sigma}+0_A\oplus i\Omega_B\geq0$, that is, $(U_{A}\otimes U_{B})\rho(U_{A}\otimes U_{B})^{\dagger}\in{\mathcal US}_{A\rightarrow B}^{G}$.
\hfill$\Box$\fi

{\bf Theorem 3.5.} {\it Let $\Phi(K,M,\bar{d})$ be an $(m+n)$-mode Gaussian channel  acting on the CV system described by $H_{A}\otimes H_{B}$. If $M+0_A\oplus i\Omega_B-K(0_A\oplus i\Omega_B)K^{\rm T}\geq0$, then $\Phi({\mathcal US}_{A\rightarrow B}^{G})\subseteq {\mathcal US}_{A\rightarrow B}^{G}$.}

 {  {The set of local Gaussian channels mentioned in Theorem 3.3 constitutes a particular instance of Theorem 3.5.}}

  A  Gaussian channel $\Phi(K,M,\bar{d})$ is called steering breaking if it sends every Gaussian state into Gaussian unsteerable state. Now it is also clear that

 {\bf Proposition 3.6.}  A  Gaussian channel $\Phi(K,M,\bar{d})$ is steering breaking if $M+0_A\oplus i\Omega_B-iK(\Omega_A\oplus\Omega_{B})K^{\rm T}\geq 0$.

Note that, if $\Phi$ is an $n$-mode Gaussian channel, then there exist real matrices $K$, $M=M^{\rm T}$, and a vector $\bar{d}$ such that $M+i\Omega_n-iK\Omega_nK^{\rm T}\geq 0$, and $\Gamma_{\Phi(\rho)}=K\Gamma_\rho K^{\rm T}+M$ holds for every Gaussian state $\rho$.  However, there  exists  $n$-mode channel $\Phi(K,M,\bar{d})$ which send every Gaussian state to Gaussian state but $M+i\Omega_n-iK\Omega_nK^{\rm T}\not\geq 0$ (see Example 3.7). This reveals that, in the definition of Gaussian channels, the requirement $M+i\Omega_n-iK\Omega_nK^{\rm T}\geq 0$ is only a sufficient condition for a channel $\Phi(K,M,\bar{d})$  to send every Gaussian state into a Gaussian state.

 Akin to the concept of Gaussian channels, we emphasize that,  for any $(m+n)$-mode Gaussian channel $\Phi(K,M,\bar{d})$ acting on the CV system $H_{A}\otimes H_{B}$, $M+0_A\oplus i\Omega_B-K(0_A\oplus i\Omega_B)K^{\rm T}\geq0$ is  a sufficient  condition for $\Phi$ to send Gaussian unsteerable states into Gaussian unsteerable states. This condition, though necessary in certain contexts, does not guarantee the exclusivity of this behavior across all possible Gaussian channels or states. The next example for $(1+1)$-mode Gaussian channel reveals that  this condition is not  necessary.

{\bf Example 3.7.}  There exists $(1+1)$-mode quantum channel $\Phi_1(K_1,M_1,\bar{d}_1)$ which sends every Gaussian state into Gaussian state but $M_1+i(\Omega_A\oplus \Omega_B)-iK_1(\Omega_A\oplus \Omega_B)K_1^{\rm T}\not\geq 0$. There exists $(1+1)$-mode Gaussian channel $\Phi_2(K_2,M_2,\bar{d}_2)$ which sends every Gaussian unsteerable state into Gaussian unsteerable state but $M_2+0_A\oplus i\Omega_B -K_2(0_A\oplus i\Omega_B)K_2^{\rm T}\not\geq 0$.

Now, we give the following definition.

{\bf Definition 3.8.}  Let $\Phi(K,M,\bar{d})$ be an $(m+n)$-mode Gaussian channel  acting on the CV system described by $H_{A}\otimes H_{B}$. $\Phi(K,M,\bar{d})$ is  Gaussian unsteerable  if $M+0_A\oplus i\Omega_B-K(0_A\oplus i\Omega_B)K^{\rm T}\geq0$;  $\Phi(K,M,\bar{d})$ is maximal Gaussian unsteerable if $\Phi({\mathcal US}_{A\rightarrow B}^{G})\subseteq {\mathcal US}_{A\rightarrow B}^{G}$.

By  the above definition, we have two choices for the notion of free operations:

 {\bf Choice 1}:  {\it A Gaussian channel $\Phi$ is free for Gaussian steering  if $\Phi$ is Gaussian unsteerable;}

{\bf Choice 2}: {\it A Gaussian channel $\Phi$ is free for Gaussian steering  if $\Phi$ is maximal Gaussian unsteerable.}

Obviously,  the choice 1 will be more convenient when dealing with the tasks in quantum information processing as the Gaussian unsteerable channels are easily constructed, recognized and applied. But
 be aware of that, by this choice, the set of free operations is a proper (but large) subset of the set of all Gaussian channels   transform Gaussian unsteerable states into Gaussian unsteerable states, that is, the maximal Gaussian unsteerable  channels.

\subsection{The measures of Gaussian steering}
\if false In the quantum resource theory, once the sets of free states and free operations  is established, it is necessary to show the existence of a  measure $\mu$ which is a nonlinear functional from the space of  Gaussian states into the non-negative real numbers   satisfying two fundamental rules: $\mu(\rho)=0$ if and only if the Gaussian state $\rho$ is a free state and $\mu(\Phi(\rho))\leq\mu(\rho)$ for all free operations $\Phi$. In this section, we propose, for Gaussian steering, three Gaussian steering measures based on the distances between two Gaussian states and discuss their properties.\fi

In the framework of quantum resource theory, upon defining the sets of free states and free operations, it becomes essential to establish the existence of a measure $\mu$, a nonlinear functional that maps the space of Gaussian states into the non-negative real numbers. This measure must fulfill two fundamental criteria: firstly, the faithfulness, that is, $\mu(\rho)=0$ if and only if the Gaussian state $\rho$ is a free state; secondly,  the non-increasing property under free operations, that is, for any free operation $\Phi$, the inequality $\mu(\Phi(\rho))\leq\mu(\rho)$ must hold for all Gaussian states $\rho$. In this section, for both choices of free operations, we introduce three measures specifically tailored for assessing Gaussian steering, each grounded in the notion of distances between Gaussian states. We then proceed to analyze and discuss the properties of these proposed measures.

{\bf Definition 3.9.}  For any $(m+n)$-mode Gaussian state $\rho$, define
$${\mathcal N}_{1}(\rho)=\mathop{\rm inf}_{\sigma\in{\mathcal US}_{A\rightarrow B}^{G}}S(\rho||\sigma),$$
$${\mathcal N}_{2}(\rho)=1-\mathop{\rm sup}_{\sigma\in{\mathcal US}_{A\rightarrow B}^{G}}\mathcal{A}(\rho, \sigma),$$
$${\mathcal N}_{3}(\rho)=1-\mathop{\rm sup}_{\sigma\in{\mathcal US}_{A\rightarrow B}^{G}}\mathcal{F}(\rho, \sigma),$$
where $S(\rho||\sigma)={\rm Tr}(\rho{\rm log}\rho)-{\rm Tr}(\rho{\rm log}\sigma)$, $\mathcal{A}(\rho, \sigma)={\rm Tr}\sqrt{\rho}\sqrt{\sigma}$ and $\mathcal{F}(\rho, \sigma)=({\rm Tr}\sqrt{\sqrt{\rho}\sigma\sqrt{\rho}})^{2}$ are the relative entropy, affinity and Uhlmann fidelity between $\rho$ and $\sigma$, respectively. Moreover, the infimum and supremum are taken over all Gaussian unsteerable states $\sigma$ from $A$ to $B$ by Alice's Gaussian measurements.

In the following, we will prove that ${\mathcal N}_{j}$ is a well-defined measure of the Gaussian steering for each $j=1,2,3$ by Theorems 3.10-3.12.

{\bf Theorem 3.10.} (Faithfulness) {{\it   For any $(m+n)$-mode Gaussian state $\rho$ and each $j\in\{1,2,3\}$, we have ${\mathcal N}_{j}(\rho)\geq 0$}. Moreover, ${\mathcal N}_{j}(\rho)=0$ if and only if $\rho\in{\mathcal US}_{A\rightarrow B}^{G}$.}

{\bf Theorem 3.11.} {\it For each  $j\in\{1,2,3\}$,   ${\mathcal N}_{j}((U_A\otimes U_B)\rho
(U_A\otimes U_B)^{\dagger})={\mathcal N}_{j}(\rho)$ holds for all $(m+n)$-mode Gaussian states $\rho$ and all local
Gaussian unitary operators $U_A\in{\mathcal B}(H_A)$ and $U_B\in{\mathcal B}(H_B)$.}

The next result reveals that, for each $j\in\{1,2,3\}$, $\mathcal N_j$ is  non-increasing under any maximal Gaussian unsteerable channels and consequently, under any Gaussian unsteerable channels.

{\bf Theorem 3.12.}  (Non-increasing property under maximal Gaussian unsteerable channels) {\it  If $\Phi $ is a maximal Gaussian unsteerable channel acting on an $(m+n)$-mode CV system $H_{A}\otimes H_{B}$, then for any $(m+n)$-mode Gaussian state $\rho$, we have ${\mathcal N}_{j}(\Phi(\rho))\leq{\mathcal N}_{j}(\rho)$ for each $j\in\{1,2,3\}$.}

In summary, whether we choose Gaussian unsteerable channels as the free operations  or maximal Gaussian unsteerable channels as free operations,  $\mathcal N_j$, $j=1,2,3$, are true Gaussian steering measures and  therefore, we have shown that {\it the Gaussian steering is really a quantum resource}.

\if false Generally speaking, ${\mathcal N}_{j}$ is difficult to be calculated for each $j=1,2,3$. However, by \cite{CP}, if either $\rho$ or $\sigma$ is in a pure state, then $\mathcal{F}(\rho, \sigma)= {\rm Tr}(\rho\sigma)$, which makes the calculation significantly simpler. Therefore, the following result gives a computation formula of ${\mathcal N}_{3}$ for any $(1+1)$-mode Gaussian pure states.\fi

We have no analytic formula of $\mathcal N_j$ for general $(m+n)$-mode Gaussian states. However,  the following example provides
an upper bound of $\mathcal  N_3(\rho)$ at some special Gaussian states.

{\bf Example 3.13.}  For any $(1+1)$-mode Gaussian pure state $\rho$ with covariance matrix $$\Gamma_\rho=\left(\begin{array}{cccc}r&0&\sqrt{r^{2}-1}&0\\0&r&0&-\sqrt{r^{2}-1}\\\sqrt{r^{2}-1}&0&r&0\\

0&-\sqrt{r^{2}-1}&0&r\end{array}\right),$$
where $r\geq1$, we have ${\mathcal N}_{3}(\rho)\leq1-\frac{4}{r+3}$.

\section{Two computable quantifications of Gaussian steering}\label{sec:4}

We have seen that Gaussian steering from $A$ to $B$ by Alice's Gaussian measurements is a quantum resource. However, the Gaussian steering measures $\mathcal N_j$, $j=1,2,3$, { proposed in Section \ref{sec:3} are difficult to be computed}. This is because, besides the difficulties of calculating the values $S(\rho||\sigma)$, $\mathcal{A}(\rho, \sigma)$ and $\mathcal{F}(\rho, \sigma)$, the optimization procedure inf or sup is also involved, which makes the Gaussian steering is not easy to be applied in real scenarios. Therefore, it is useful to quantify Gaussian steering so that the quantifications are easily accessible, though they may not be the true Gaussian steering measures.

In this section, we present two computable quantifications of the Gaussian steering, and discuss their properties.

{\bf Definition 4.1.} For any $(m+n)$-mode Gaussian state $\rho$ with covariance matrix $\Gamma_\rho$, define
$$ {\mathcal J}_1(\rho)=\frac{\|\Gamma_\rho +0_A\oplus i\Omega_B\|_1}{{\rm Tr}(\Gamma_\rho)}-1, \eqno(4.1)
$$
and
$$ {\mathcal J}_2(\rho)={\|\Gamma_\rho +0_A\oplus i\Omega_B\|_1}-{{\rm Tr}(\Gamma_\rho)}, \eqno(4.2)
$$
where $\|\cdot\|_1$ stands for the trace-norm, that is, $\|F\|_1={\rm Tr}((F^\dag F)^{\frac{1}{2}})$.

We show that ${\mathcal J}_j$ is a quantification of the Gaussian steering from $A$ to $B$ for each $j=1,2$. To do this, {we need a recent mathematical result appeared in \cite{EGL} and a proof of it is given in \cite{LL}}. Denote by $\mathcal{T}(H)$ the trace-class of all operators $T$ with $\|T\|_1<\infty$ on a Hilbert space $H$.

{\bf Lemma 4.2.} (\cite{EGL,LL}) {\it Let $H$ be a complex or real Hilbert space, $T\in\mathcal{T}(H)$. Then $\|T\|_1={\rm Tr}(T) $ if and only if $T\geq 0$.}

By applying this lemma, we can show that

{\bf Theorem 4.3.} (Faithfulness) {\it For $j\in\{1,2\}$, ${\mathcal J}_j(\rho)\geq 0$ holds for all $(m+n)$-mode Gaussian state $\rho$. Moreover, ${\mathcal J_j}(\rho)=0$ if and only if $\rho\in{\mathcal US}_{A\rightarrow B}^{G}$.}

{\bf Theorem 4.4.} {\it Let $\rho_{1}$ and $\rho_{2}$ be $(m+n)$-mode Gaussian states
with covariance matrices $\Gamma_{\rho_{1}}$ and $\Gamma_{\rho_{2}}$, respectively.
Let $\Gamma_{\rho}=p_{1}\Gamma_{\rho_{1}}+p_{2}\Gamma_{\rho_{2}}$ and
$\rho$ be the $(m+n)$-mode Gaussian state with covariance matrix $\Gamma_{\rho}$, where
$p_{1},p_{2}\geq0$, $p_{1}+p_{2}=1$. Then
$${\mathcal J}_1(\rho)\leq{\mathcal J}_1(\rho_{1})+{\mathcal J}_1(\rho_{2})+1,$$
and
$${\mathcal J}_2(\rho)\leq
p_{1}{\mathcal J}_2(\rho_{1})+p_{2}{\mathcal J}_2(\rho_{2}).$$}

The above result focuses on the study of convex combinations of covariance matrices
of Gaussian states since a convex combination of Gaussian states is not necessarily a
Gaussian state, while a convex combination of covariance matrices of Gaussian states
must be the covariance matrix of a Gaussian state.

In the following, we discuss the non-increasing property of ${\mathcal J}_j(\rho)$ under certain Gaussian unsteerable channels for any $(m+n)$-mode Gaussian state $\rho$.

{\bf Theorem 4.5.} {\it For any $(m+n)$-mode Gaussian state $\rho$ and any Gaussian channels $\Phi_{A}(K_{A},M_{A},\bar{d}_{A})$ and $\Phi_{B}(K_{B},M_{B},\bar{d}_{B})$ acting on the subsystems $H_{A}$ and $H_{B}$, respectively. If $K_{A}$, $K_{B}$ are the orthogonal matrices and $K_{B}$ is a symplectic matrix, then ${\mathcal J}_j((\Phi_{A}\otimes \Phi_{B})\rho)\leq {\mathcal J}_j(\rho)$, $j=1,2.$}

A small generalization of Theorem 4.5 is true.

{\bf Theorem 4.6.} {\it For any $(m+n)$-mode Gaussian state $\rho$ and any Gaussian channel $\Phi=\Phi(K,M,\bar{d})$ with $K=K_A\oplus K_B$,  if $K_{A}$, $K_{B}$ are the orthogonal matrices and $K_{B}$ is a symplectic matrix, then ${\mathcal J}_j(\Phi(\rho))\leq {\mathcal J}_j(\rho)$, $j=1,2$.}

 The classical noise channel is presented as an important example of the Gaussian noise channel \cite{WHT}. Recall that an $n$-mode  $\Phi(K, M, \bar{d})$ is a classical noise channel if $K=I_{2n}$ and $\bar{d}=0$.
By Theorem 4.6, we see that ${\mathcal J}_j$, $j=1,2$, is non-increasing under the classical noise channel for any $(m+n)$-mode Gaussian state $\rho$.

Theorems 4.3-4.6 reveal that both ${\mathcal J}_1$ and $\mathcal J_2$ are  faithful quantifications of the Gaussian steering from $A$ to $B$ which are non-increasing under certain Gaussian unsteerable channels. {Specifically, we have discovered that Theorem 4.5 is useful when preparing samples for machine learning purposes.} Each ${\mathcal J}_j(\rho)$ can be calculated at any $(m+n)$-mode Gaussian state $\rho$. Furthermore, the computation of $\mathcal J_j$ is solely dependent on the covariance matrices of Gaussian states, thereby circumventing the need for conventional optimization procedures. This significantly diminishes the computational complexity involved, rendering these quantifications highly practical and efficient tools for analyzing Gaussian steering.

However, neither $\mathcal J_1$ nor $\mathcal J_2$
  is a definitive Gaussian steering measure, as they fail to exhibit the property of being non-increasing under the entire spectrum of Gaussian  unsteerable channels,  as the following example shows.

{\bf Example 4.7.}  For $j=1,2$, there exists a $(1+1)$-mode local Gaussian channel $\Phi=\Phi_{A}\otimes \Phi_{B}$ such that ${\mathcal J}_j((\Phi_{A}\otimes \Phi_{B})\rho)> {\mathcal J}_j(\rho)$ for some $(1+1)$-mode Gaussian state $\rho$.

Next, we present some analytical expressions for ${\mathcal J}_j$ at the $(m+n)$-mode Gaussian pure state, specifically for a standard form of the covariance matrix, and also for a special class of $(1+1)$-mode Gaussian states. Both cases are parameterized by the elements of the covariance matrix.

 According to \cite{GJJ}, the covariance matrix $\Gamma$ of any $(m+n)$-mode pure Gaussian state can always be brought into the phase-space Schmidt form $\Gamma_{S}$ (via a local symplectic operation $S=S_{m}\oplus S_{n}$ ), where
$$
\Gamma_{S}=\left(\begin{array}{cc}A& C\\
C^{\rm T} & B\end{array}\right),\eqno(4.3)
$$
with $A=\mathop\oplus\limits_{k=1}^{m}\left(\begin{array}{cc}\gamma_{k}& 0\\
0&\gamma_{k}\end{array}\right)$, $B=\mathop\oplus\limits_{k=1}^{m}\left(\begin{array}{cc}\gamma_{k}& 0\\
0&\gamma_{k}\end{array}\right)\oplus I_{2(n-m)}$ and
$$C=\left(\begin{array}{ccccccccc}D_{1}&\cdot\cdot\cdot&0&0&\cdot\cdot\cdot&0\\
\vdots&\ddots&\vdots&\vdots&\cdot\cdot\cdot&\vdots\\
0&\cdot\cdot\cdot&D_{m}&0&\cdot\cdot\cdot&0\\
\end{array}\right)_{2m\times2n}$$
if $m\leq n$, and
$A=\mathop\oplus\limits_{k=1}^{n}\left(\begin{array}{cc}\gamma_{k}& 0\\
0&\gamma_{k}\end{array}\right)\oplus I_{2(m-n)}$, $B=\mathop\oplus\limits_{k=1}^{n}\left(\begin{array}{cc}\gamma_{k}& 0\\
0&\gamma_{k}\end{array}\right)$ and
$$C=\left(\begin{array}{ccccccccc}D_{1}&\cdot\cdot\cdot&0\\
\vdots&\ddots&\vdots\\
0&\cdot\cdot\cdot&D_{n}\\
0& \cdot\cdot\cdot & 0 \\
\vdots &\ddots &\vdots \\
0 &\cdot \cdot \cdot & 0
\end{array}\right)_{2m\times2n}$$
if $m>n$.

In matrix $C$, $0$ corresponds to the $2\times2$ zero matrix and $$D_{k}=\left(\begin{array}{cc}\sqrt{\gamma_{k}^{2}-1}& 0\\
0&-\sqrt{\gamma_{k}^{2}-1}\end{array}\right), k=1,2,\cdots,\min\{m,n\}.$$
Here $\gamma_{k}\geq1$ $(k=1,2,\cdots,\min\{m,n\})$  is the mode
(A$_{k}$)-(B$_{k}$) mixing factor. We call $\Gamma_{S}$  the phase-space Schmidt form
of $\Gamma$.

 {\bf Theorem 4.8.} {\it For $(m+n)$-mode pure Gaussian state $\rho$ with covariance
matrix $\Gamma_{S}$  in Eq.(4.3),  let $\gamma_{k}\geq1$ $(k=1,2,\ldots,\min\{m,n\})$  be  the mode
(A$_{k}$)-(B$_{k})$ mixing factor
of $\Gamma_{S}$.
Then we have
$${\mathcal J}_1(\rho)=\frac{\sum_{k=1}^{\min\{m,n\}}(1+2\gamma_{k}+\sqrt{4\gamma_{k}^{2}-3})+2|n-m|}
{\sum_{k=1}^{\min\{m,n\}}(4\gamma_{k})+2|n-m|}-1$$
and
$${\mathcal J}_2(\rho)=\sum_{k=1}^{\min\{m,n\}}(1-2\gamma_{k}+\sqrt{4\gamma_{k}^{2}-3}).$$
 Moreover, a pure Gaussian state $\rho$ is unsteerable from A to B if and only if $\gamma_{k}=1$ for all $k=1,2,\ldots,\min\{m,n\}$.}

Among the $(1+1)$-mode Gaussian state $\rho$ with covariance matrix $\Gamma_{0}$ in Eq.(2.3), there exist some special states that are experimentally important. If $c=d$, the Gaussian state is a mixed thermal state. When $c=-d$, the Gaussian state is a squeezed thermal state. The following theorem gives the expressions of ${\mathcal J}_j$, $j=1,2$,
for these two classes of $(1+1)$-mode Gaussian states.

{\bf Theorem 4.9.} {\it For a $(1+1)$-mode Gaussian state $\rho$ with covariance matrix $\Gamma_0$ in Eq.(2.3) with $c=|d|$, we have
$${\mathcal J}_1(\rho)={\rm max}\{0, \frac{1+a+b+\sqrt{(a-b+1)^{2}+4c^{2}}}{2(a+b)}-1\}$$
and
$${\mathcal J}_2(\rho)={\rm max}\{0, 1+\sqrt{(a-b+1)^{2}+4c^{2}}-(a+b)\}.$$
Moreover, if $a(b-1)-c^{2}\geq0$, then $\rho\in{\mathcal US}_{A\rightarrow B}^{G}$.}

\section{Comparing $\mathcal J_2$ with Gaussian steering measure $\mathcal N_3$}\label{sec:5}

${\mathcal J}_j$ ($j=1,2$), as a quantification of  Gaussian steering,  can be used to detect the Gaussian steering for any $(m+n)$-mode Gaussian state. Moreover, as ${\mathcal J}_j$ is easily calculated,  it is advantageous for  detecting Gaussian steering and can replace the role of Gaussian steering measure in many tasks of quantum information processes.

In this section, we compare ${\mathcal J}_2$ with the  Gaussian steering measure $\mathcal{N}_{3}$ at  $(1+1)$-mode Gaussian pure state as we have drawn an upper bound for this situation (see Example 3.13).

For $(1+1)$-mode Gaussian pure state $\rho$ with covariance matrix $$\Gamma_\rho=\left(\begin{array}{cccc}r&0&\sqrt{r^{2}-1}&0\\0&r&0&-\sqrt{r^{2}-1}\\\sqrt{r^{2}-1}&0&r&0\\
0&-\sqrt{r^{2}-1}&0&r\end{array}\right), \eqno(5.1)$$
where $r\geq1$. By Example 3.13 and Theorem 4.8, we have
$${\mathcal N}_3(\rho)\leq z(\rho)=1-\frac{4}{r+3},\eqno(5.2)$$
$${\mathcal J}_2(\rho)=1-2r+\sqrt{4r^{2}-3}.\eqno(5.3)$$

Note that $1-\frac{4}{r+3}\leq1-2r+\sqrt{4r^{2}-3}$ as $1-2r+\sqrt{4r^2-3}-1+\frac{4}{r+3}=\frac{(r+3)\sqrt{4r^2-3}+(4-2r^2-6r)}{r+3}\geq0$ for $r\geq1$. This is because for $r\geq1$, {we have} $r+3>0$, $(r+3)\sqrt{4r^2-3}\geq0$, $4-2r^2-6r\leq0$ and $[(r+3)\sqrt{4r^2-3}]^{2}-(4-2r^2-6r)^{2}=13r^2+30r-43\geq0$. Hence, we always have
$$\mathcal N_3(\rho)\leq z(\rho)\leq \mathcal J_2(\rho) \eqno(5.4)$$
for all $(1+1)$-mode pure Gaussian states with the covariance matrices having the form as in Eq.(5.1), and ``$=$" holds if and only if $r=1$.

\begin{figure}[h]
  \centering
\includegraphics[width=7cm,height=5.8cm]{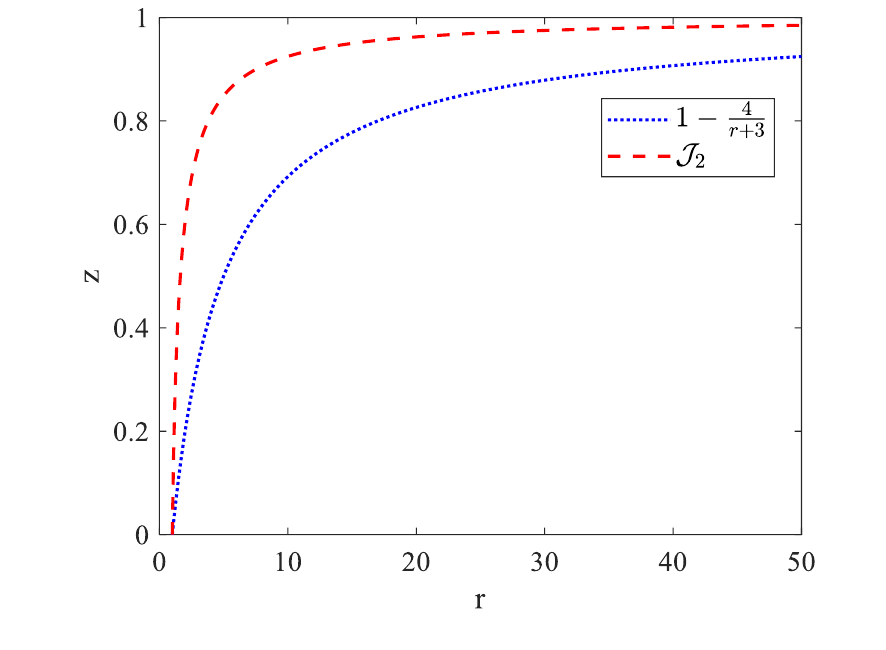}
\caption{\quad   The {dotted blue}  line  is $z=1-\frac{4}{r+3},$ and the { dashed red} line expresses $ z={\mathcal J}_2(\rho)$ when the $(1+1)$-mode Gaussian pure state $\rho$ regarded as a function of the parameter $r$.}
\end{figure}

 Fig.1 demonstrates the behaviors of $z=1-\frac{4}{r+3}$ and ${\mathcal J}_2$ at $(1+1)$-mode Gaussian pure state when regarding ${\mathcal J}_2$ as function of $r$. At $r=1$, ${\mathcal J}_2(\rho)=1-\frac{4}{r+3}=\mathcal{N}_{3}(\rho)=0$, which is because the $(1+1)$-mode Gaussian pure state $\rho$ is a Gaussian unsteerable  state at $r=1$. For $r>1$, ${\mathcal J}_2(\rho)>1-\frac{4}{r+3}\geq\mathcal{N}_{3}(\rho)$, which reveals that, as an upper bound of $\mathcal N_3$, $\mathcal J_2$ has advantage when detect the steering of $\rho$, especially for those with $r$ near 1.


\section{Behaviour of Gaussian steering in Markovian environments}\label{sec:6}

An application of accessible quantifications of Gaussian steering lies in elucidating its behavior within the context of system evolutionary processes. As an illustration,  in this section, for $(1+1)$-mode Gaussian states, we study the behaviour of the Gaussian steering in Gaussian noise environments by applying the quantification ${\mathcal J}_2$. Furthermore, we focus on the scenarios in Markovian environments. By \cite{HPZ,ISM}, it is possible to extend the analysis to non-Markovian environments.

The dynamics of a $(1+1)$-mode quantum state $\rho$ through a (Markovian) noisy
environment is governed by the following Master equation (\cite{GKS}-\cite{OS}):
$$\dot{\rho}=\sum_{k=1}^{2}\frac{\lambda}{2}\{(N+1)\mathcal{L}[\hat{a}_k]+N\mathcal{L}[\hat{a}_k^\dag]-M^{*}\mathcal{D}[\hat{a}_k]-M\mathcal{D}[\hat{a}_k^\dag]\}\rho,$$
where $\hat{a}_k^\dag$ and $\hat{a}_k$ are the creation and annihilation
operators in the $k$th mode. $\mathcal{L}[O]\rho=2O\rho O^{^\dag}-O^{^\dag}O\rho-\rho O^{^\dag}O$ and $\mathcal{D}[O]\rho=2O\rho O-OO\rho-\rho OO$ are Lindblad superoperators. $\lambda$ is the overall damping rate, $N\in{\mathbb R}$ and $M\in{\mathbb C}$ represent the effective number of photons and the squeezing parameter of the bath, respectively with $|M|^{2}\leq N(N+1)$.

Time evolution imposed by the master equation preserves the Gaussian character of the state $\rho(t)$ and the
covariance matrix at time $t$ is given by (\cite{OS}-\cite{SPF})
$$\Gamma_{\rho(t)}=e^{-\lambda t}\Gamma_{\rho(0)}+(1-e^{-\lambda t})\Gamma_{\rho(\infty)} \eqno(6.1)$$
with
\begin{eqnarray}
&&\Gamma_{\rho(\infty)}\nonumber\\&=&2\left(\begin{array}{cccc}\frac{1}{2}+L_{+}&M_{I}&0&0\\M_{I}&\frac{1}{2}+L_{-}&0&0\\0&0&\frac{1}{2}+L_{+}&M_{I}\\
0&0&M_{I}&\frac{1}{2}+L_{-}\end{array}\right).\nonumber \end{eqnarray}
Here, $L_{\pm}=N\pm M_{R}$, $N=n_{th}({\rm cosh}^{2}(R)+{\rm sinh}^{2}(R))+{\rm sinh}^{2}R$, $M=-(2n_{th}+1){\rm cosh}(R){\rm sinh}(R)e^{i\phi}$, $n_{th}$ is the thermal photon number, $R$ is the squeezing parameter of the bath, and $\phi$ is the squeezing phase. $M_{R}$ and $M_{I}$ represent
the real and the imaginary part of $M$, respectively. By Theorem 4.4, one sees that
 $${\mathcal J}_2(\rho(t))\leq e^{-\lambda t}{\mathcal J}_2(\rho(0))+(1-e^{-\lambda t}){\mathcal J}_2(\rho(\infty)). \eqno(6.2)$$

In order to study the behaviour of Gaussian steering  in Markovian environments, we need to detect the Gaussian steering contained in $\rho(t)$. To do this, we use $\mathcal J_2$ to measure $\rho(t)$, and consider the behaviour of $\mathcal J_2(\rho(t))$.  By Eq.(6.1),
$$\begin{array}{rl} \mathcal J_2(\rho(t))= & \|e^{-\lambda t}\Gamma_{\rho(0)}+(1-e^{-\lambda t})\Gamma_{\rho(\infty)} +0_{A}\oplus i\Omega_{B} \|_1 \\
& -{\rm Tr}(e^{-\lambda t}\Gamma_{\rho(0)}+(1-e^{-\lambda t})\Gamma_{\rho(\infty)}).\end{array}\eqno(6.3)$$

Assume that the initial Gaussian state $\rho(0)$ is a squeezed vacuum state, of which the covariance matrix is
$$\Gamma_{\rho(0)}=\left(\begin{array}{cccc}{\rm cosh}2r&0&{\rm sinh}2r&0\\0&{\rm cosh}2r&0&-{\rm sinh}2r\\{\rm sinh}2r&0&{\rm cosh}2r&0\\
0&-{\rm sinh}2r&0&{\rm cosh}2r\end{array}\right),$$
where $r$ is the squeezing parameter of the state.  By  Theorem 4.9, ${\mathcal J}_2(\rho(0))=0$ if $r=0$, and
$${\mathcal J}_2(\rho(0))=1+\sqrt{4{\rm cosh}^{2}2r-3}-2{\rm cosh}2r $$
otherwise.

\begin{figure}[h]
  \centering
\includegraphics[width=7cm,height=12cm]{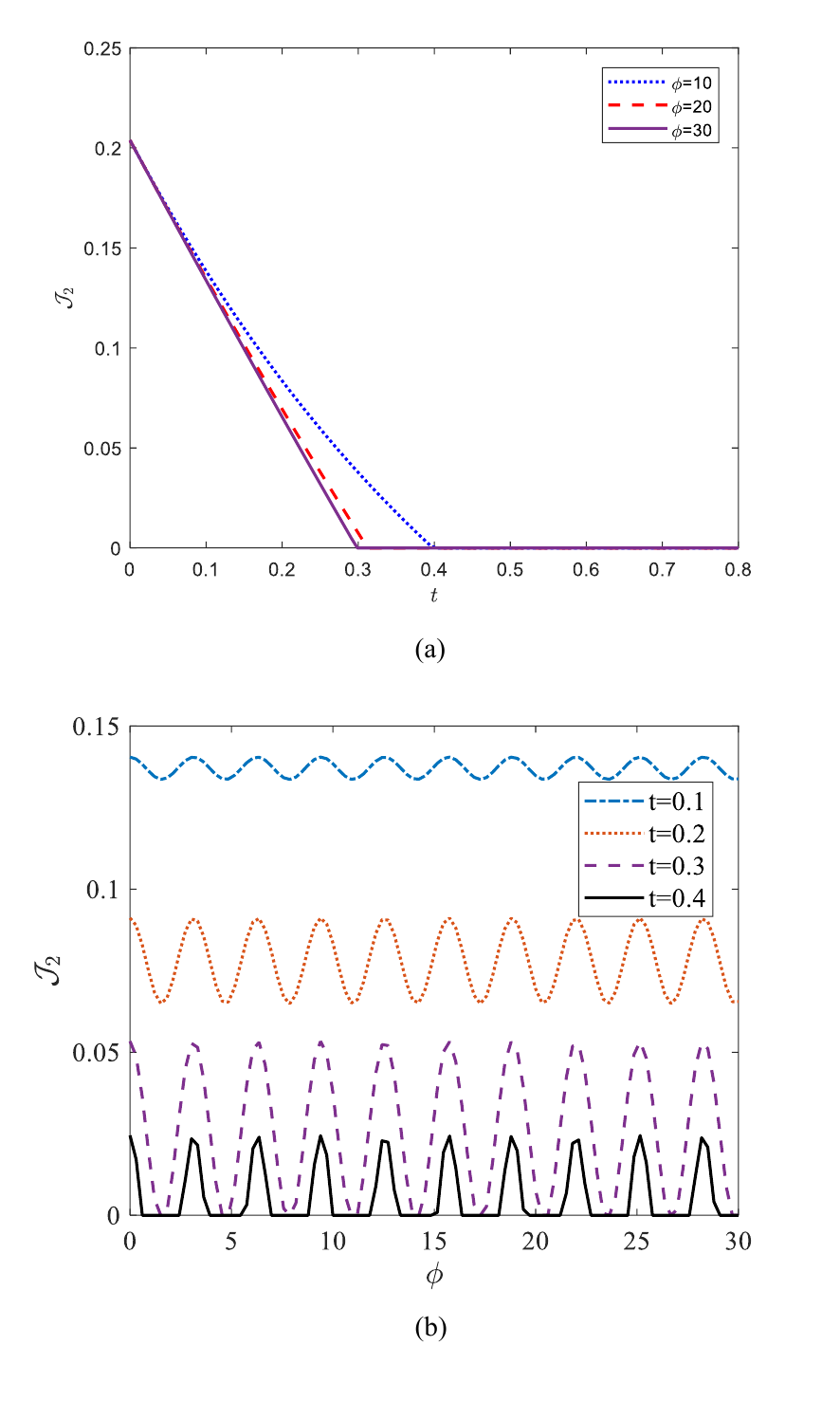}
\caption{\quad  Behaviour of    ${\mathcal J}_2(\rho(t))$ {with $\rho(0)$ the squeezed vacuum state of $r=1$}, as a function of the parameters $t$ and $\phi$ for fixed $n_{th}=0$, $R=1$ and $\lambda=0.1$. The figure (a) shows ${\mathcal J}_2$ as a function of the parameter $t$ for fixed $\phi=10$, $\phi=20$ and $\phi=30$, while the figure (b) shows ${\mathcal J}_2$ as a function of the parameter $\phi$ for fixed $t=0.1$, $t=0.2$, $t=0.3$ and $t=0.4$.}
\end{figure}

\begin{figure*}[]\label{figure5}
\center
\subfigure []
{\includegraphics[width=7cm,height=5.5cm]{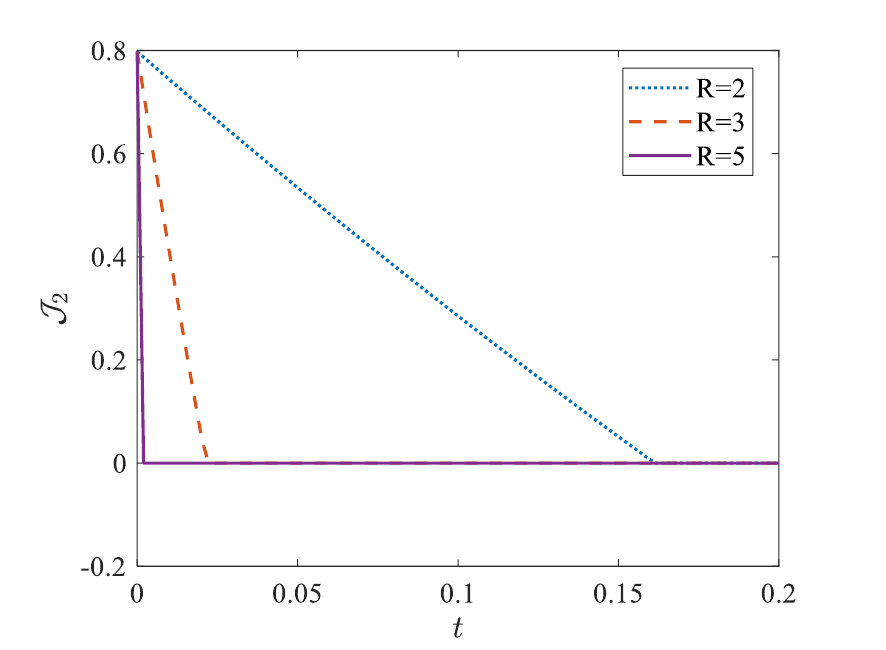}}
\subfigure []
{\includegraphics[width=7cm,height=5.5cm]{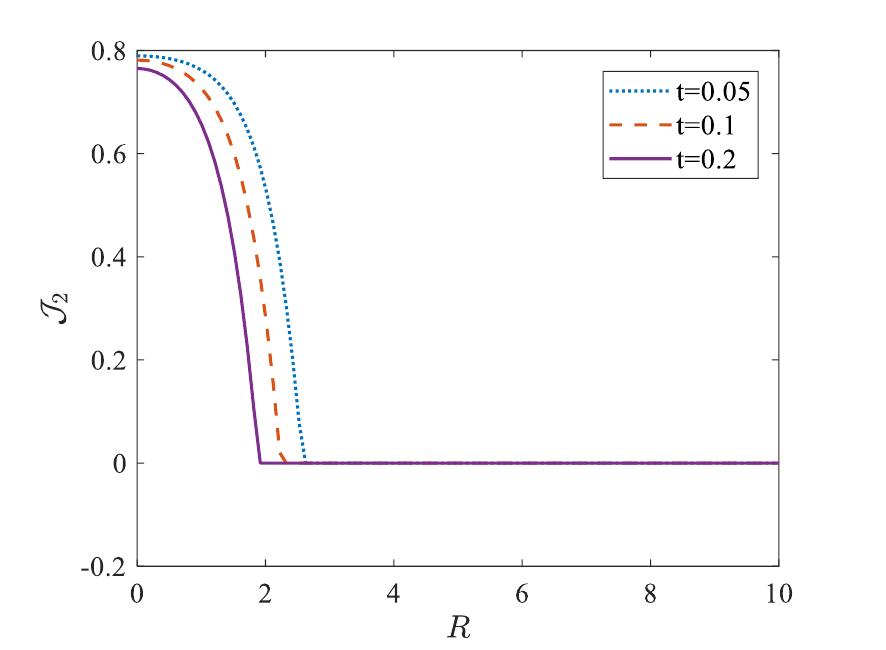}}
  \caption{Behaviour of  ${\mathcal J}_2(\rho(t))$ {with $\rho(0)$ the squeezed vacuum state of $r=1$}, as a function of the parameters $t$ and $R$ for fixed $n_{th}=0$, $\phi=0$ and $\lambda=0.1$. The figure (a) shows ${\mathcal J}_2$ as a function of the parameter $t$ for fixed $R=2$, $R=3$ and $R=5$, while the figure (b) shows ${\mathcal J}_2$ as a function of the parameter $R$ for fixed $t=0.05$, $t=0.1$ and $t=0.2$.}
\end{figure*}

\begin{figure*}[]\label{figure6}
\center
\subfigure []
{\includegraphics[width=7cm,height=5.5cm]{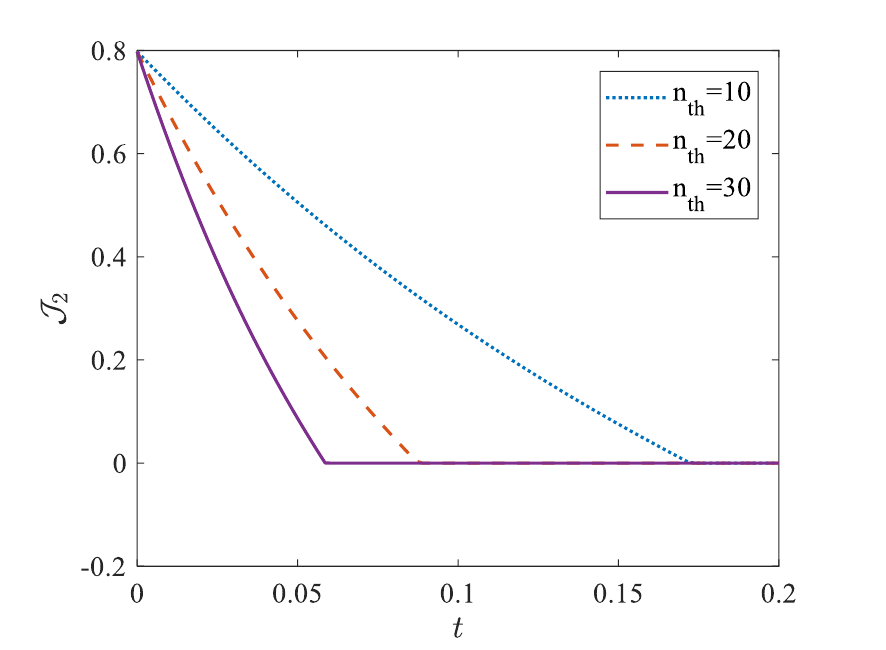}}
\subfigure []
{\includegraphics[width=7cm,height=5.5cm]{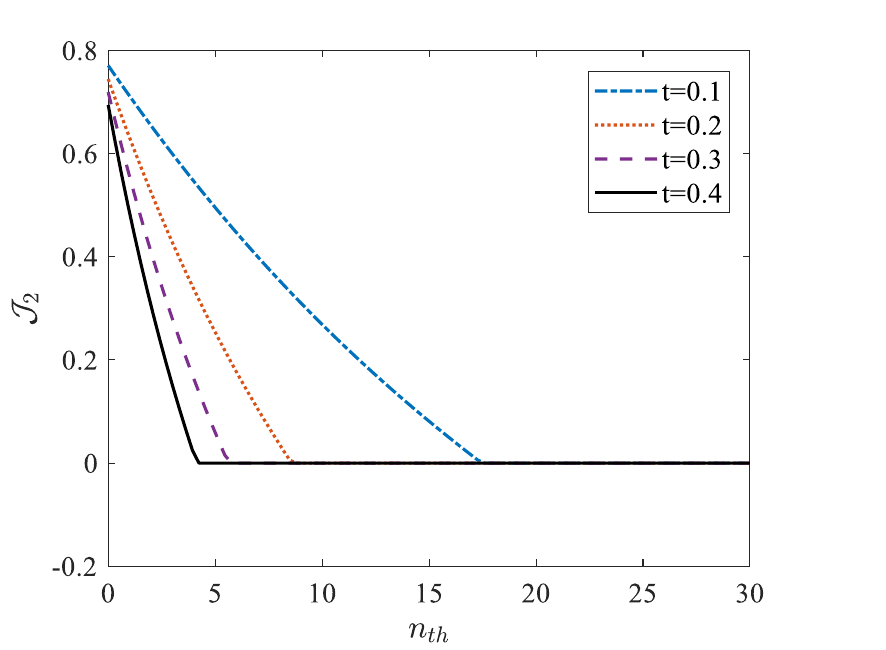}}

\caption{\quad  Behaviour of  ${\mathcal J}_2(\rho(t))$ {with $\rho(0)$ the squeezed vacuum state of $r=1$}, as a function of the parameters $t$ and $n_{th}$ for fixed $R=0.5$, $\phi=0$ and $\lambda=0.1$. The figure (a) shows ${\mathcal J}_2$ as a function of the parameter $t$ for fixed $n_{th}=10$, $n_{th}=20$ and $n_{th}=30$, while the figure (b) shows ${\mathcal J}_2$ as a function of the parameter $n_{th}$ for fixed $t=0.1$, $t=0.2$, $t=0.3$ and $t=0.4$}
\centering
\end{figure*}

In Fig.2, when the parameters $n_{th}$, $R$ and $\lambda$ in $\Gamma_{\rho(\infty)}$ are fixed, the magnitude of ${\mathcal J}_2(\rho(t))$, with $r$ set to 1,  exhibits a monotonic decreasing behavior over time, gracefully decaying towards 0.
Fig.2(a) reveals that, as $\phi$ increases, the value of ${\mathcal J}_2(\rho(t))$  declines more rapidly. While Fig.2(b) shows that, at time $t$,  ${\mathcal J}_2(\rho(t))$ exhibits an oscillatory behaviour as a function of parameter $\phi$, and as time increases, the magnitude of the oscillations first increases and then decreases.

Fig.3 demonstrates the behaviour of   ${\mathcal J}_2(\rho(t))$ {with $\rho(0)$ the squeezed vacuum state of $r=1$}, as a function of the parameters $t$ and $R$ for fixed $n_{th}$, $\phi$, $\lambda$. From Fig.3(a), as $R$ increases, the speed of ${\mathcal J}_2$ {reducing to 0} increases.
Furthermore,   Fig.3(b) reveals that, for fixed $t$, the  ${\mathcal J}_2$ is also decreasing and decays to 0 as a function of $R$.

Fig.4 illustrates the case that   ${\mathcal J}_2(\rho(t))$ {with $\rho(0)$ the squeezed vacuum state of $r=1$} is regarded as a function of the parameters $t$ and $n_{th}$ for fixed $R$, $\phi$, $\lambda$. From Fig.4(a), we notice that, as $n_{th}$ increases, the rate of decline in ${\mathcal J}_2$ becomes more pronounced. Fig.4(b) shows the similar behaviour of  ${\mathcal J}_2$ when regard it as a function of $n_{th}$.

\section{Conclusion and discussion}\label{sec:7}

\if false The existent quantum resource theory for Gaussian steering from party $A$ to party $B$ by A's Gaussian measurements has  defects and is far from perfect. The main defect lies in the lack of understanding of the structure of unsteerable Gaussian channels, which leads to an overly narrow choice of free operations. One of the main contributions of this work is its exhaustive and in-depth examination of the structural complexities of Gaussian unsteerable channels, offering a more comprehensive and nuanced understanding. And then, define free operations as all Gaussian unsteerable channels, thereby establishing a complete quantum resource theory for Gaussian steering. Due to the complexity of the  definitions, the known genuine Gaussian steering measures  cannot be effectively calculated, which greatly limits the application of Gaussian steering as a quantum resource in quantum information processes. Another contribution of this work is to propose  two quantifications $\mathcal J_1$ and $\mathcal J_2$ of Gaussian steering, which only involve the covariance matrices of Gaussian states and do not involve any optimization process, and thus can be directly calculated at any $(m+n)$-mode Gaussian state. Although $\mathcal J_1$ and $\mathcal J_2$  are not Gaussian steering measures, they have some good properties. Especially, for any $(1+1)$-mode Gaussian pure state,  $\mathcal J_2$ is an upper bound of the Gaussian steering measure $\mathcal N_3$, of which the definition is based on the  Uhlmann fidelity. Applying  $\mathcal J_2$,  the behaviour of Gaussian steering is discussed for a special class of $(1+1)$-mode Gaussian states under Gaussian noise channels, which uncovers the intriguing phenomenon of rapid decay in quantum steering, shedding light on the dynamics of non-classical correlations under certain conditions.\fi

The current quantum resource theory for Gaussian steering, from $A$ to $B$ by Alice's Gaussian measurements, is flawed and incomplete. Its primary shortcoming stems from an inadequate comprehension of the architecture of  Gaussian channels transforming Gaussian unsteerable states into Gaussian unsteerable states, resulting in a restricted selection of free operations. One of the main contributions of this work is  to provide a thoroughly and deeply examining  the structural intricacies of such Gaussian  channels, fostering a more exhaustive and nuanced understanding. Consequently, the concepts of Gaussian unsteerable channels and maximal Gaussian unsteerable channels are introduced, both of them may be chosen as the free operations. Thereby we formulate a comprehensive quantum resource theory for Gaussian steering as several genuine Gaussian steering measures are proposed.

Note that, a Gaussian channel $\Phi=\Phi(K,M,\bar{d})$ is defined as a Gaussian unsteerable channel if $M+0_A\oplus i\Omega_B-K(0_A\oplus i\Omega_B)K^{\rm T}\geq 0$, a condition that is easily constructed, recognized and applicable.
It is well known that an $n$-mode Gaussian channel  $\Phi$ can be determined by real matrices $K$, $M=M^{\rm T}$ satisfying $M+i\Omega_n-iK\Omega_n K^{\rm T}\geq 0$ and a vector $\bar{d}$ via $\Gamma_{\Phi(\rho)}=K\Gamma_\rho K^{\rm T} +M$, ${\bf d}_{\Phi(\rho)}=K{\bf d}_\rho +\bar{d}$, and thus one may write $\Phi=\Phi(K,M,\bar{d})$. There is some misunderstanding when we say that a trace preserving completely positive superoperator $\Phi$ of $n$-mode CV system is a Gaussian channel if it sends $n$-mode Gaussian states into Gaussian states.  In fact, there exists quantum channel $\Phi(K,M,\bar{d})$ transforming Gaussian states into Gaussian states but   $M+i\Omega_n-iK\Omega_n K^{\rm T}\not\geq 0$.
Akin to the concept of Gaussian channels, we emphasize that,  for any $(m+n)$-mode Gaussian channel $\Phi=\Phi(K,M,\bar{d})$ acting on the CV system $H_{A}\otimes H_{B}$, $M+0_A\oplus i\Omega_B-K(0_A\oplus i\Omega_B)K^{\rm T}\geq0$ is also a sufficient  condition for $\Phi$ to send Gaussian unsteerable states into Gaussian unsteerable states. This condition does not necessary for $\Phi$ to be Gaussian unsteerable. Also note that, every two-way GLOCC channel is Gaussian unsteerable. Therefore,
 the set of all Gaussian unsteerable channels  encompass a proper but large subset of the set of all  maximal Gaussian unsteerable channels.

Due to the intricacy of the definitions, existing genuine Gaussian steering measures are challenging to compute, significantly limiting their application in quantum information processes as a quantum resource. In response, this work introduces two quantifications, $\mathcal J_1$ and $\mathcal J_2$,  for Gaussian steering. These quantifications solely rely on the covariance matrices of Gaussian states, eliminating the need for optimization, enabling direct computation for any $(m+n)$-mode Gaussian state.
While $\mathcal J_1$ and $\mathcal J_2$  are not  genuine  measures of Gaussian steering, they possess some desirable properties, {which may be useful in many  scenarios, for instance, when preparing samples for machine learning purposes}. Notably, for certain class of $(1+1)$-mode Gaussian pure states, $\mathcal J_2$ serves as an upper bound for the genuine Gaussian steering measure $\mathcal N_3$, which is based on the Uhlmann fidelity. By utilizing $\mathcal J_2$,  we investigate  the behaviour of Gaussian steering in a particular class of (1+1)-mode Gaussian states in Markovian environments, revealing the  phenomenon of rapid decay in quantum steering. This insight sheds light on the dynamics of non-classical correlations under certain conditions.

{\bf Acknowledgments}
This work was supported by the National Natural Science Foundation
 of China (Grant Nos. 12071336, 12171290, 12271394)

{\bf InterestConflict} {The authors declare that they have no conflict of interest.}

\section*{appendix}
\section*{A: Proofs of the results in Section \ref{sec:3}}

In this appendix section we present all proofs of the results and examples appeared in {  Section \ref{sec:3}}.

{\bf Proof of Theorem 3.1.} We have to show that for any given  sequence $\{\rho_{k}\}_{k=1}^\infty$ of Gaussian unsteerable states, if $\lim_{k}\|\rho_{k}-\rho\|_{1}=0$ for some trace class operator $\rho$, then we have  $\rho\in{\mathcal US}_{A\rightarrow B}^{G}$. Denote by $\Gamma_{k}$ and ${\mathbf d}_{k}$ the covariance matrix and the mean vector of $\rho_{k}$, respectively; then $\Gamma_{k}+0_{A}\oplus i\Omega_B\geq0$ for each $k$.

By \cite{LRX}, $\Gamma_{k}$ and ${\mathbf d}_{k}$ are bounded sequences, that is, there exists $M\in \mathbb{R}$ such that $\max_k\{\|\Gamma_{k}\|_{\infty}, |{\mathbf d}_{k}|_{2}\}\leq M$, where $\|\cdot\|_{\infty}$ is the operator norm, and $|\cdot|_{2}$ is the Euclidean norm for vectors. Since $\Gamma_{k}$ and ${\mathbf d}_{k}$ live in finite-dimensional spaces, they will admit  simultaneously convergent subsequences $\Gamma_{k_{s}}$ and ${\mathbf d}_{k_{s}}$, that is, there exists $\Gamma\in \mathcal M_{2(m+n)}(\mathbb R)$ and $\mathbf{d}\in \mathbb R^{2(m+n)}$ such that $\Gamma_{k_{s}}\rightarrow\Gamma$, ${\mathbf d}_{k_{s}}\rightarrow{\mathbf d}$. It is clear that $\displaystyle\lim_{s\rightarrow\infty}\|\rho(\Gamma_{k_{s}}, {\mathbf d}_{k_{s}})-\rho\|_{1}=0$ and $\Gamma_{k_{s}}+0_{A}\oplus i\Omega_B\geq0$ for each $s$. Moreover, since the set of covariance matrices of the Gaussian unsteerable states is topologically closed, we have $\Gamma+0_{A}\oplus i\Omega_B\geq0$.

By Eq.(2.1), $${\rm Tr}(\rho(\Gamma_{k_{s}}, {\mathbf d}_{k_{s}}) W(z))=\exp[-\frac{1}{4}z^{\rm T}\Gamma_{k_{s}} z+i{\mathbf d}_{k_{s}}^{\rm T}z].$$
Clearly,
$$\lim_{s\rightarrow\infty}\exp[-\frac{1}{4}z^{\rm T}\Gamma_{k_{s}} z+i{\mathbf d}_{k_{s}}^{\rm T}z]=\exp[-\frac{1}{4}z^{\rm T}\Gamma z+i{\mathbf d}^{\rm T}z].$$
On the  other hand, since the convergence of the sequence of states is in trace norm and $W(z)$ is a bounded operator, one gets
$$\lim_{s\rightarrow\infty}{\rm Tr}(\rho(\Gamma_{k_{s}}, {\mathbf d}_{k_{s}}) W(z))={\rm Tr}(\rho  W(z))=\chi_{\rho}(z).$$
Therefore, we have $\chi_{\rho}(z)=\exp[-\frac{1}{4}z^{\rm T}\Gamma z+i{\mathbf d}^{\rm T}z]$, that is, the covariance matrix of $\rho$ is $\Gamma$ and $\rho$ is Gaussian.
Since $\Gamma+0_{A}\oplus i\Omega_B\geq0$,   $\rho\in{\mathcal US}_{A\rightarrow B}^{G}$.
\hfill$\Box$

{\bf Proof of Theorem 3.2.} Since $V$ is covariance matrix of a Gaussian unsteerable state from $A$ to $B$, by Eq.(2.4), $V+0_{A}\oplus i\Omega_B\geq0$. And $W\geq V$ entails $W+0_{A}\oplus i\Omega_B\geq V+0_{A}\oplus i\Omega_B\geq0$. Consequently, $W$ is also the covariance matrix of some Gaussian unsteerable state from $A$ to $B$.
\hfill$\Box$

{\bf Proof of Theorem 3.3.} Denote by $\Gamma_\rho=\left(\begin{array}{cc}A& C\\
C^{\rm T} & B\end{array}\right)$ the covariance matrix of $\rho$.  Since $\rho\in{\mathcal US}_{A\rightarrow B}^{G}$, we have $\Gamma_\rho +0_A\oplus i\Omega_B\geq 0$, that is,

$$\left(\begin{array}{cc}A& C\\
C^{\rm T} & B+i\Omega_B\end{array}\right)\geq 0.\eqno(A.1)$$

By Eq.(3.1), the covariance matrix $\Gamma_{\sigma}$ of $\sigma=(\Phi_{A}\otimes I_{B})\rho$ is
$$\Gamma_{\sigma}=\left(\begin{array}{cc}K_{A}& 0\\
0 & I_{B}\end{array}\right)\left(\begin{array}{cc}A& C\\
C^{\rm T} & B\end{array}\right)\left(\begin{array}{cc}K_{A}^{{\rm T}}& 0\\
0 & I_{B}\end{array}\right)+\left(\begin{array}{cc}M_{A}& 0\\
0 & 0\end{array}\right).$$

By Eq.(A.1) and $M_{A}\geq0$, it is clear that
\begin{eqnarray}
&&\Gamma_{\sigma}+0_A\oplus i\Omega_B\nonumber\\&=&\left(\begin{array}{cc}K_{A}& 0\\
0 & I_{B}\end{array}\right)\left(\begin{array}{cc}A& C\\
C^{\rm T} & B\end{array}\right)\left(\begin{array}{cc}K_{A}^{{\rm T}}& 0\\
0 & I_{B}\end{array}\right)\nonumber\\&&+\left(\begin{array}{cc}M_{A}& 0\\
0 & 0\end{array}\right)+\left(\begin{array}{cc}0& 0\\
0 & i\Omega_B\end{array}\right)\nonumber\\&=&\left(\begin{array}{cc}K_{A}& 0\\
0 & I_{B}\end{array}\right)\left(\begin{array}{cc}A& C\\
C^{\rm T} & B+i\Omega_B\end{array}\right)\left(\begin{array}{cc}K_{A}^{{\rm T}}& 0\\
0 & I_{B}\end{array}\right)\nonumber\\&&+\left(\begin{array}{cc}M_{A}& 0\\
0 & 0\end{array}\right)\geq0\nonumber
\end{eqnarray}
Consequently, $\sigma=(\Phi_{A}\otimes I_{B})\rho\in{\mathcal US}_{A\rightarrow B}^{G}$.

Similarly, one can check that $\Gamma_{(I_{A}\otimes \Phi_{B})\rho}+0_A\oplus i\Omega_B\geq0$, that is, $(I_{A}\otimes \Phi_{B})\rho\in{\mathcal US}_{A\rightarrow B}^{G}$. There is only one point where $M_{B}+i\Omega_{B}-iK_{B}\Omega_{B} K_{B}^{\rm T}\geq0$ is used in the argument of checking $\Gamma_{(I_{A}\otimes \Phi_{B})\rho}+0_A\oplus i\Omega_B\geq0$.

Now, for any $(m+n)$-mode Gaussian state $\rho\in{\mathcal US}_{A\rightarrow B}^{G}$, we see that $(\Phi_{A}\otimes\Phi_{B})\rho=(I_A\otimes \Phi_B)(\Phi_A\otimes I_B)\rho\in{\mathcal US}_{A\rightarrow B}^{G}$. \hfill$\Box$

{\bf Proof of Theorem 3.5.} For any $(m+n)$-mode unsteerable Gaussian state $\rho$ with the covariance matrix  $\Gamma_\rho$, we have  $\Gamma_\rho+0_A\oplus i\Omega_B\geq0$. Thus $K(\Gamma_\rho+0_A\oplus i\Omega_B)K^{\rm T}\geq0$ holds  for any $K\in {\mathcal M}_{2(m+n)}(\mathbb R)$. As the covariance matrix of $\Phi(\rho)$ is $\Gamma_{\Phi(\rho)}=K\Gamma_\rho K^{\rm T}+M$, the assumption $M+0_A\oplus i\Omega_B-K(0_A\oplus i\Omega_B)K^{\rm T}\geq0$ ensures that
$$\begin{array}{rl} &\Gamma_{\Phi(\rho)}+0_A\oplus i\Omega_B\\=&K\Gamma_\rho K^{\rm T}+M+0_A\oplus i\Omega_B \\
= & K(\Gamma_\rho+0_A\oplus i\Omega_B)K^{\rm T}+M\\&+0_A\oplus i\Omega_B-K(0_A\oplus i\Omega_B)K^{\rm T}\\ \geq&0,\end{array}$$
which implies that $\Phi(\rho)\in{\mathcal US}_{A\rightarrow B}^{G}$.
\hfill$\Box$

{\bf Proof of Proposition 3.6.} Obvious.

{\bf Proof of Example 3.7.} The proof is numerical.

{Let $\Phi_{1}(K_{1},M_{1},\bar{d}_{1})$ be a $(1+1)$-mode quantum channel with $$K_{1}=\left(\begin{array}{cccc}-353.124&-257.135&-43.9143&61.7854\\-517.650&829.322&-7.6448&-42.2212\\339.674&-933.669&-14.2333&68.6708\\
-465.469&-377.374&-53.5241&56.0476\end{array}\right),$$
and
\begin{eqnarray}
&&M_{1}=\nonumber\\&&\left(\begin{array}{cccc}1.17\times10^{7}&2.33\times10^{6}&-5.83\times10^{5}&-3.59\times10^{6}\\2.33\times10^{6}&1.23\times10^{6}&2.39\times10^{5}&-2.36\times10^{6}\\-5.83\times10^{5}&2.39\times10^{5}&1.44\times10^{7}&1.02\times10^{7}\\
-3.59\times10^{6}&-2.36\times10^{6}&1.02\times10^{7}&1.32\times10^{7}\end{array}\right).\nonumber \end{eqnarray}
Then  $M_{1}=M_{1}^{\rm T}\geq0$ and $M_{1}+i(\Omega_{A}\oplus\Omega_B)-iK_{1}(\Omega_{A}\oplus\Omega_B) K_{1}^{\rm T}\not\geq0$. However, for randomly generating 100000 samples of the covariance matrices $\{\Gamma_\rho\}$ of (1+1)-mode Gaussian states, we consistently observe $K_{1}\Gamma_\rho K_{1}^{\rm T}+M_{1}+i(\Omega_{A}\oplus\Omega_B)\geq0$.
Therefore, $\Phi(K_{1},M_{1},\bar{d}_{1})$ is largely a Gaussian channel for any $\bar{d}_{1}$. So the condition
$M_{1}+i(\Omega_{A}\oplus\Omega_B)-iK_{1}(\Omega_{A}\oplus\Omega_B) K_{1}^{\rm T}\geq0$  may not be necessary for $\Phi_{1}(K_{1},M_{1},\bar{d}_{1})$ to send Gaussian states into Gaussian states.}

{Let $\Phi_{2}(K_{2},M_{2},\bar{d}_{2})$ be a $(1+1)$-mode Gaussian channel with
\begin{eqnarray}
&&K_{2}\nonumber\\&=&\left(\begin{array}{cccc}0.89540737&0.13270765&0.28588588&0.75217447\\0.90747409&0.75837409&0.0462667&0.6361504\\0.58844177&0.94277329&0.64957331&0.11012731\\
0.22886036&0.85541575&0.96036917&0.96621468\end{array}\right)\nonumber \end{eqnarray}
and
\begin{eqnarray}
&&M_{2}\nonumber\\&=&\left(\begin{array}{cccc}1.7219939&0.6023585&1.25044133&0.74670078\\0.6023585&0.90434614&0.83432618&0.12961425\\1.25044133&0.83432618&2.15765071&0.39996861\\
0.74670078&0.12961425&0.39996861&0.607965\end{array}\right).\nonumber \end{eqnarray}
Then  $M_{2}=M_{2}^{\rm T}\geq0$, $M_{2}+i(\Omega_{A}\oplus\Omega_B)-iK_{2}(\Omega_{A}\oplus\Omega_B) K_{2}^{\rm T}\geq0$ but $M_{2}+0_A\oplus i\Omega_B-K_{2}(0_A\oplus i\Omega_B)K_{2}^{\rm T}\not\geq0$. However, for randomly generating 100000 samples of the covariance matrices $\{\Gamma_\rho\}$ of (1+1)-mode Gaussian unsteerable states, we observe that  we always have $K_{2}\Gamma_\rho K_{2}^{\rm T}+M_{2}+0_A\oplus i\Omega_B\geq0$.
Therefore, $\Phi(K_{2},M_{2},\bar{d}_{2})$ is largely a Gaussian unsteerable channel for any $\bar{d}_{2}$. So the condition
$M_{2}+0_A\oplus i\Omega_B-K_{2}(0_A\oplus i\Omega_B)K_{2}^{\rm T}\geq0$ in Theorem 3.5 may not be necessary for a Gaussian  channel $\Phi_{2}(K_{2},M_{2},\bar{d}_{2})$ to send Gaussian unsteerable states into Gaussian unsteerable states.}\hfill$\Box$

{\bf Proof of Theorem 3.10.} It is clear that ${\mathcal N}_{j}(\rho)\geq0$, $j=1,2,3$,  since for   any two states $\rho$ and $\sigma$, we have $S(\rho||\sigma)\geq0$, $0\leq\mathcal{A}(\rho, \sigma)\leq1$ and $0\leq\mathcal{F}(\rho, \sigma)\leq1$.

We only give the proof for the case of $j=1$. The cases $j=2,3$ are checked similarly.

By Definition 3.9, ${\mathcal N}_{1}(\rho)=0$
if and only if \if false there exists some $\sigma_{0}$  belongs to the closure of ${\mathcal US}_{A\rightarrow B}^{G}$, such that
$S(\rho\|\sigma_{0})=0$.\fi $\rho\in$ the $\|\cdot\|_1$-closure of ${\mathcal US}_{A\rightarrow B}^{G}$, and in turn, by Theorem 3.1, if and only if $\rho\in{\mathcal US}_{A\rightarrow B}^{G}$. \if false Note that $S(\rho\|\sigma_{0})=0$ if and only if
$\rho=\sigma_{0}\in \mathcal{US}_{A\rightarrow B}^{G}$. Consequently,
it is clear that ${\mathcal N}_{1}(\rho)=0$ if and only if $\rho\in
\mathcal{US}_{A\rightarrow B}^{G}$.\fi
\hfill$\Box$

{\bf Proof of Theorem 3.11.} Since
$$S((U_A\otimes U_B)\rho
(U_A\otimes U_B)^{\dagger}||(U_A\otimes U_B)\sigma
(U_A\otimes U_B)^{\dagger})=S(\rho||\sigma),$$
$$\mathcal{A}((U_A\otimes U_B)\rho
(U_A\otimes U_B)^{\dagger}, (U_A\otimes U_B)\sigma
(U_A\otimes U_B)^{\dagger})=\mathcal{A}(\rho, \sigma),$$
$$\mathcal{F}((U_A\otimes U_B)\rho
(U_A\otimes U_B)^{\dagger}, (U_A\otimes U_B)\sigma
(U_A\otimes U_B)^{\dagger})=\mathcal{F}(\rho, \sigma),$$
and by Corollary 3.4 one can easily check that ${\mathcal N}_{j}((U_A\otimes U_B)\rho
(U_A\otimes U_B)^{\dagger})={\mathcal N}_{j}(\rho)$.
\hfill$\Box$

{\bf Proof of Theorem 3.12.} Since $\Phi$ is a maximal Gaussian unsteerable channel,  $\Phi(\sigma)\in\mathcal{US}_{A\rightarrow B}^{G}$ holds   for all $\sigma\in\mathcal{US}_{A\rightarrow B}^{G}$.

Note that $S(\Psi(\rho)||\Psi(\sigma))\leq S(\rho||\sigma)$ is true for any quantum channel $\Psi$ and any states $\rho,\sigma$. Thus we have
\begin{eqnarray}
{\mathcal N}_{1}(\Phi(\rho))&=&\mathop{\rm inf}_{\sigma\in{\mathcal US}_{A\rightarrow B}^{G}}S(\Phi(\rho)||\sigma)\nonumber\\&\leq&
\mathop{\rm inf}_{\sigma\in{\mathcal US}_{A\rightarrow B}^{G}}S(\Phi(\rho)||\Phi(\sigma))\nonumber\\&\leq&\mathop{\rm inf}_{\sigma\in{\mathcal US}_{A\rightarrow B}^{G}}S(\rho||\sigma)\nonumber\\&=&{\mathcal N}_{1}(\rho).\nonumber
\end{eqnarray}

Let $\mathcal{M}=\mathcal{A}$ or $ \mathcal{F}$. Then $\mathcal{M}(\Psi(\rho), \Psi(\sigma))\geq\mathcal{M}(\rho, \sigma)$ holds for any quantum channel $\Psi$ and any states $\rho,\sigma$.  Applying this property to $j\in\{2,3\}$,  one has
\begin{eqnarray}
{\mathcal N}_{j}(\Phi(\rho))&=&1-\mathop{\rm sup}_{\sigma\in{\mathcal US}_{A\rightarrow B}^{G}}\mathcal{M}(\Phi(\rho), \sigma)\nonumber\\&\leq&
1-\mathop{\rm sup}_{\sigma\in{\mathcal US}_{A\rightarrow B}^{G}}\mathcal{M}(\Phi(\rho), \Phi(\sigma))\nonumber\\&\leq&1-\mathop{\rm sup}_{\sigma\in{\mathcal US}_{A\rightarrow B}^{G}}\mathcal{M}(\rho, \sigma)\nonumber\\&=&{\mathcal N}_{j}(\rho).\nonumber
\end{eqnarray}
\if false
In summary, for $j=\{1,2,3\}$, ${\mathcal N}_{j}(\Phi(\rho))\leq{\mathcal N}_{j}(\rho)$.\fi
\hfill$\Box$

{\bf Proof of Example 3.13.} In fact, by \cite{KLRA}, if $(1+1)$-mode Gaussian state $\sigma\in{\mathcal US}_{A\rightarrow B}^{G}$ has covariance matrix $\Gamma_{\sigma}$ as in Eq.(2.3), then $(ab-c^{2})(ab-d^{2})\geq a^{2}$. Since $\rho$ is a Gaussian pure state, $\mathcal{F}(\rho, \sigma)= {\rm Tr}(\rho\sigma)$. By \cite{PAT}, for any $(1+1)$-mode Gaussian state $\sigma$, ${\rm Tr}(\rho\sigma)=\frac{4}{\sqrt{{\rm det}(\Gamma_{\rho}+\Gamma_{\sigma})}}$. Consequently,
\begin{eqnarray}
&&{\mathcal N}_{3}(\rho)\nonumber\\&=&1-\mathop{\rm sup}_{\sigma\in{\mathcal US}_{A\rightarrow B}^{G}}\mathcal{F}(\rho, \sigma)\nonumber\\&\leq&1-\mathop{\rm max}\{\frac{4}{\sqrt{[(r+a)(r+b)-(\sqrt{r^{2}-1}+c)^{2}]}}\nonumber\\&&\times\frac{1}{[(r+a)(r+b)-(-\sqrt{r^{2}-1}+d)^{2}]}\}\nonumber,
\end{eqnarray}
where the inequality arises from the fact that we only consider the supremum over Gaussian unsteerable states that possess covariance matrices of standard form. Here the maximum value is taken over all $a$, $b$, $c$, $d$ satisfying $a\geq1$, $b\geq1$, $a(ab-c^{2})-b\geq0$, $b(ab-d^{2})-a\geq0$, $(ab-c^{2})(ab-d^{2})+1-a^{2}-b^{2}-2cd\geq0$ and $(ab-c^{2})(ab-d^{2})\geq a^{2}$. It follows from  calculation that ${\mathcal N}_{3}(\rho)\leq1-\frac{4}{r+3}$.
\hfill$\Box$\\

\section*{B: proofs of the results in Section \ref{sec:4}}

In the appendix section we provide all proofs of the results and examples in {  Section
\ref{sec:4}}.

{\bf Proof of Theorem 4.3.} Assume $j=1$. Denote by $\Gamma_\rho$ the covariance matrix of $\rho$. Since $\|\Gamma_\rho +0_A\oplus i\Omega_B\|_1\geq {\rm Tr}(\Gamma_\rho +0_A\oplus i\Omega_B)={\rm Tr}(\Gamma_\rho)>0$, by Definition 4.1, it is clear that $ {\mathcal J}_1(\rho)=\frac{\|\Gamma_\rho +0_A\oplus i\Omega_B\|_1}{{\rm Tr}(\Gamma_\rho)}-1\geq 0$. If $\rho\in{\mathcal US}_{A\rightarrow B}^{G}$, then $\Gamma_\rho +0_A\oplus i\Omega_B\geq 0$ and hence
$\|\Gamma_\rho +0_A\oplus i\Omega_B\|_1={\rm Tr}(\Gamma_\rho+0_A\oplus i\Omega_B) ={\rm Tr}(\Gamma_\rho)$, which ensures that $\mathcal J_1(\rho)=0$. Conversely, $\mathcal J_1(\rho)=0$ implies that $\|\Gamma_\rho +0_A\oplus i\Omega_B\|_1={\rm Tr}(\Gamma_\rho)={\rm Tr}(\Gamma_\rho+0_A\oplus i\Omega_B) $. By Lemma 4.2, we see that $\Gamma_\rho +0_A\oplus i\Omega_B\geq 0$ and hence $\rho\in{\mathcal US}_{A\rightarrow B}^{G}$.

The proof for the case $j=2$ is similar. \hfill$\Box$

{\bf Proof of Theorem 4.4.} Since ${\rm Tr}(\Gamma_{\rho})={\rm Tr}(p_{1}\Gamma_{\rho_{1}}+p_{2}\Gamma_{\rho_{2}})=p_{1}{\rm Tr}(\Gamma_{\rho_{1}})+p_{2}{\rm Tr}(\Gamma_{\rho_{2}})$, one gets
\begin{eqnarray}
&&\|\Gamma_{\rho}+0_A\oplus i\Omega_B\|_1\nonumber\\&=&\|p_{1}\Gamma_{\rho_{1}}+p_{1}(0_A\oplus i\Omega_B)+p_{2}\Gamma_{\rho_{2}}+p_{2}(0_A\oplus i\Omega_B)\|_1\nonumber\\&\leq&\|p_{1}\Gamma_{\rho_{1}}+p_{1}(0_A\oplus i\Omega_B)\|_1+\|p_{2}\Gamma_{\rho_{2}}+p_{2}(0_A\oplus i\Omega_B)\|_1\nonumber\\&=&p_{1}\|\Gamma_{\rho_{1}}+0_A\oplus i\Omega_B\|_1+p_{2}\|\Gamma_{\rho_{2}}+0_A\oplus i\Omega_B\|_1.\nonumber
\end{eqnarray}
Thus,
\begin{eqnarray}
&&{\mathcal J}_1(\rho)\nonumber\\&=&\frac{\|\Gamma_\rho +0_A\oplus i\Omega_B\|_1}{{\rm Tr}(\Gamma_\rho)}-1\nonumber\\&\leq&\frac{p_{1}\|\Gamma_{\rho_{1}}+0_A\oplus i\Omega_B\|_1+p_{2}\|\Gamma_{\rho_{2}}+0_A\oplus i\Omega_B\|_1}{p_{1}{\rm Tr}(\Gamma_{\rho_{1}})+p_{2}{\rm Tr}(\Gamma_{\rho_{2}})}-1\nonumber\\&\leq&\frac{p_{1}\|\Gamma_{\rho_{1}}+0_A\oplus i\Omega_B\|_1}{p_{1}{\rm Tr}(\Gamma_{\rho_{1}})}+\frac{p_{2}\|\Gamma_{\rho_{2}}+0_A\oplus i\Omega_B\|_1}{p_{2}{\rm Tr}(\Gamma_{\rho_{2}})}-1\nonumber\\&=&{\mathcal J}_1(\rho_{1})+{\mathcal J}_1(\rho_{2})+1\nonumber
\end{eqnarray}
and
\begin{eqnarray}
&&{\mathcal J}_2(\rho)\nonumber\\&=&{\|\Gamma_\rho +0_A\oplus i\Omega_B\|_1}-{{\rm Tr}(\Gamma_\rho)}\nonumber\\&\leq&p_{1}\|\Gamma_{\rho_{1}}+0_A\oplus i\Omega_B\|_1+p_{2}\|\Gamma_{\rho_{2}}+0_A\oplus i\Omega_B\|_1\nonumber\\&&-p_{1}{\rm Tr}(\Gamma_{\rho_{1}})-p_{2}{\rm Tr}(\Gamma_{\rho_{2}})\nonumber\\&=&
p_{1}{\mathcal J}_2(\rho_{1})+p_{2}{\mathcal J}_2(\rho_{2}).\nonumber
\end{eqnarray}

\hfill$\Box$

{\bf Proof of Theorem 4.5.} Denote by $\Gamma_{\rho}$ the covariance matrix of $\rho$.

We first prove that ${\mathcal J}_j((\Phi_{A}\otimes I_{B})\rho)\leq {\mathcal J}_j(\rho).$ Since $M_{A}\geq0$, then $M_{A}\oplus 0_{B}\geq0$. By
Lemma 4.2, we see that $\|M_{A}\oplus 0_{B}\|_1={\rm Tr}(M_{A}\oplus 0_{B})$.

Assume $j=1$.  Note that $\frac{b+c}{a+c}\leq\frac{b}{a}$ whenever $b\geq a>0$ and $c\geq 0$. By
Definition 4.1 and $K_{A}^{{\rm T}}K_{A}=I_{A}$, we have
\begin{eqnarray}
&&{\mathcal J}_1((\Phi_{A}\otimes I_{B})\rho)\nonumber\\&=&\frac{\|\Gamma_{(\Phi_{A}\otimes I_{B})\rho} +0_A\oplus i\Omega_B\|_1}{{\rm Tr}[\Gamma_{(\Phi_{A}\otimes I_{B})\rho}]}-1\nonumber\\&=&\frac{\|(K_{A}\oplus I_{B})\Gamma_\rho (K_{A}\oplus I_{B})^{\rm T}+M_{A}\oplus 0_{B}+0_A\oplus i\Omega_B\|_1}{{\rm Tr}[(K_{A}\oplus I_{B})\Gamma_\rho (K_{A}\oplus I_{B})^{\rm T}+M_{A}\oplus 0_{B}]}-1\nonumber\\&\leq&\frac{\|(K_{A}\oplus I_{B})\Gamma_\rho (K_{A}\oplus I_{B})^{\rm T}+0_A\oplus i\Omega_B\|_1}{{\rm Tr}[(K_{A}\oplus I_{B})\Gamma_\rho (K_{A}\oplus I_{B})^{\rm T}]+{\rm Tr}(M_{A}\oplus 0_{B})}\nonumber\\&&+\frac{\|M_{A}\oplus 0_{B}\|_1}{{\rm Tr}[(K_{A}\oplus I_{B})\Gamma_\rho (K_{A}\oplus I_{B})^{\rm T}]+{\rm Tr}(M_{A}\oplus 0_{B})}-1\nonumber\\&\leq&\frac{\|(K_{A}\oplus I_{B})\Gamma_\rho (K_{A}\oplus I_{B})^{\rm T}+0_A\oplus i\Omega_B\|_1}{{\rm Tr}[(K_{A}\oplus I_{B})\Gamma_\rho (K_{A}\oplus I_{B})^{\rm T}]}-1\nonumber\\&=&\frac{\|(K_{A}\oplus I_{B})(\Gamma_\rho+0_A\oplus i\Omega_B)(K_{A}\oplus I_{B})^{\rm T}\|_1}{{\rm Tr}[(K_{A}\oplus I_{B})\Gamma_\rho (K_{A}\oplus I_{B})^{\rm T}]}-1\nonumber\\&=&\frac{\|\Gamma_\rho+0_A\oplus i\Omega_B\|_1}{{\rm Tr}(\Gamma_\rho)}-1={\mathcal J}_1(\rho).\nonumber
\end{eqnarray}

Assume $j=2$. By
Definition 4.1 and $K_{A}^{{\rm T}}K_{A}=I_{A}$, we have
\begin{eqnarray}
&&{\mathcal J}_2((\Phi_{A}\otimes I_{B})\rho)\nonumber\\&=&{\|\Gamma_{(\Phi_{A}\otimes I_{B})\rho} +0_A\oplus i\Omega_B\|_1}-{{\rm Tr}(\Gamma_{(\Phi_{A}\otimes I_{B})\rho})}\nonumber\\&=&\|(K_{A}\oplus I_{B})\Gamma_\rho (K_{A}\oplus I_{B})^{\rm T}+M_{A}\oplus 0_{B}+0_A\oplus i\Omega_B\|_1\nonumber\\&&-{\rm Tr}[(K_{A}\oplus I_{B})\Gamma_\rho (K_{A}\oplus I_{B})^{\rm T}+M_{A}\oplus 0_{B}]\nonumber\\&\leq&\|(K_{A}\oplus I_{B})\Gamma_\rho (K_{A}\oplus I_{B})^{\rm T}+0_A\oplus i\Omega_B\|_1+\|M_{A}\oplus 0_{B}\|_1\nonumber\\&&-{\rm Tr}[(K_{A}\oplus I_{B})\Gamma_\rho (K_{A}\oplus I_{B})^{\rm T}-{\rm Tr}(M_{A}\oplus 0_{B})]\nonumber\\&=&\|(K_{A}\oplus I_{B})(\Gamma_\rho +0_A\oplus i\Omega_B)(K_{A}\oplus I_{B})^{\rm T}\|_1\nonumber\\&&-{\rm Tr}[(K_{A}\oplus I_{B})\Gamma_\rho(K_{A}\oplus I_{B})^{\rm T}]\nonumber\\&=&{\|\Gamma_{\rho} +0_A\oplus i\Omega_B\|_1}-{{\rm Tr}(\Gamma_{\rho})}={\mathcal J}_2(\rho).\nonumber
\end{eqnarray}

 Next, let us show that ${\mathcal J}_j((I_{A}\otimes \Phi_{B})\rho)\leq {\mathcal J}_j(\rho)$. Since $M_{B}\geq0$, then $0_{A}\oplus M_{B}\geq0$. By
Lemma 4.2, we see that $\|0_{A}\oplus M_{B}\|_1={\rm Tr}(0_{A}\oplus M_{B})$. Then, similarly to the argument as above, we have
\begin{eqnarray}&&{\mathcal J}_1((I_{A}\otimes \Phi_{B})\rho)\nonumber\\ &\leq&\frac{\|(I_{A}\oplus K_{B})\Gamma_\rho(I_{A}\oplus K_{B})^{{\rm T}} +0_A\oplus i\Omega_B\|_1}{{\rm Tr}[(I_{A}\oplus K_{B})\Gamma_\rho(I_{A}\oplus K_{B})^{{\rm T}}]}-1\nonumber
\end{eqnarray}
and
\begin{eqnarray}&&{\mathcal J}_2((I_{A}\otimes \Phi_{B})\rho)\nonumber\\ &\leq&{\|(I_{A}\oplus K_{B})\Gamma_\rho(I_{A}\oplus K_{B})^{{\rm T}} +0_A\oplus i\Omega_B\|_1}\nonumber\\&&-{{\rm Tr}((I_{A}\oplus K_{B})\Gamma_\rho(I_{A}\oplus K_{B})^{\rm T})}.\nonumber\end{eqnarray}
Moreover, as $K_B$ is symplectic and $K_{B}^{{\rm T}}K_{B}=I_{B}$, then
\begin{eqnarray}
&&{\mathcal J}_1((I_{A}\otimes \Phi_{B})\rho)\nonumber\\ &\leq&\frac{\|(I_{A}\oplus K_{B})(\Gamma_\rho+0_A\oplus i\Omega_B)(I_{A}\oplus K_{B})^{{\rm T}}\|_1}{{\rm Tr}[(I_{A}\oplus K_{B})\Gamma_\rho(I_{A}\oplus K_{B})^{{\rm T}}]}-1\nonumber\\&=&\frac{\|\Gamma_\rho+0_A\oplus i\Omega_B\|_1}{{\rm Tr}(\Gamma_\rho)}-1\nonumber\\&=&{\mathcal J}_1(\rho)\nonumber
\end{eqnarray}
and
\begin{eqnarray}
&&{\mathcal J}_2((I_{A}\otimes \Phi_{B})\rho)\nonumber\\ &\leq&{\|(I_{A}\oplus K_{B})(\Gamma_\rho+0_A\oplus i\Omega_B)(I_{A}\oplus K_{B})^{{\rm T}}\|_1}\nonumber\\&&-{{\rm Tr}((I_{A}\oplus K_{B})\Gamma_\rho(I_{A}\oplus K_{B})^{\rm T})}\nonumber\\&=&{\|\Gamma_\rho+0_A\oplus i\Omega_B\|_1}-{{\rm Tr}(\Gamma_\rho)}\nonumber\\&=&{\mathcal J}_2(\rho).\nonumber
\end{eqnarray}

Now, combining the above inequalities just proved, one achieves
$${\mathcal J}_j((\Phi_{A}\otimes \Phi_{B})\rho)\leq {\mathcal J}_j(\rho),$$
as deired.
\hfill$\Box$

{\bf Proof of Example 4.7.} To see this, we consider $\mathcal J_2$. Let $\Phi=\Phi_A\otimes \Phi_B$ with $\Phi_{A}=\Phi_{A}(K_{A},M_{A},\bar{d}_{A})$ and $\Phi_{B}=\Phi_{B}(K_{B},M_{B},\bar{d}_{B})$ single mode Gaussian channels acting respectively on CV systems $H_A$ and $H_B$, where $K_{A}=\left(\begin{array}{cc}1&1\\0&1\end{array}\right)$, $K_{B}=\left(\begin{array}{cc}1&0\\0&1\end{array}\right)$, $M_{A}=M_{B}=0$. Let
$\rho$ be a $(1+1)$-mode Gaussian state whose covariance matrix $$\Gamma_{\rho}=\left(\begin{array}{cccc}7.84&-5&5.84&7.63\\-5&9.30&0.82&-0.71\\5.84&0.82&12.92&15.45\\
7.63&-0.71&15.45&19.01\end{array}\right).$$ By Definition 4.1, ${\mathcal J}_2(\rho)\approx0.0148$. However,
\begin{eqnarray}
&&\Gamma_{(\Phi_{A}\otimes \Phi_{B})\rho}\nonumber\\&=&\left(\begin{array}{cc}K_{A}&0\\0&K_{B}\end{array}\right)\Gamma_{\rho}\left(\begin{array}{cc}K_{A}^{\rm T}&0\\0&K_{B}^{\rm T}\end{array}\right)+\left(\begin{array}{cc}M_{A}&0\\0&M_{B}\end{array}\right)\nonumber\\
&=&\left(\begin{array}{cccc}7.14&4.30&6.66&6.92\\4.30&9.30&0.82&-0.71\\6.66&0.82&12.92&15.45\\
6.92&-0.71&15.45&19.01\end{array}\right),\nonumber
\end{eqnarray}
and hence ${\mathcal J}_2((\Phi_{A}\otimes \Phi_{B})\rho)\approx0.0152>{\mathcal J}_2(\rho)$.

Note that, $K_{A}$, $K_{B}$ are symplectic. So, this example says in fact that there exist
Gaussian unitary operators $U_{A}$, $U_{B}$,  and Gaussian state $\rho$  such that ${\mathcal J}_2((U_{A}\otimes U_{B})\rho(U_{A}^{\dag}\otimes U_{B}^{\dag}))>{\mathcal J}_2(\rho)$.\hfill$\Box$

{\bf Proof of Theorem 4.8.} Assume $m\leq n$. For any $(m+n)$-mode pure Gaussian state $\rho$ with covariance matrix $\Gamma_{S}$ in Eq.(4.3), ${\rm Tr}(\Gamma_{S})=\sum_{k=1}^{m}(4\gamma_{k})+2(n-m)$, and since $(\Gamma_{S}+0_A\oplus i\Omega_B)^{\dagger}=\Gamma_{S}+0_A\oplus i\Omega_B$,
$\|\Gamma_{S}+0_A\oplus i\Omega_B\|_1=|\lambda_{1}|+|\lambda_{2}|+...+|\lambda_{2(m+n)}|$, where $\lambda_{1},\lambda_{2},...,\lambda_{2(m+n)}$ are the eigenvalues of $\Gamma_{S}+0_A\oplus i\Omega_B$.

By calculation,  the  eigenvalues of $\Gamma_{S}+0_A\oplus i\Omega_B$ are exactly
$\underbrace{0,0,\ldots,0}_{n-m}, \underbrace{2,2,\ldots,2}_{n-m},$
$\frac{1+2\gamma_{k}+\sqrt{4\gamma_{k}^{2}-3}}{2}$, $\frac{1+2\gamma_{k}-\sqrt{4\gamma_{k}^{2}-3}}{2}$, $\frac{2\gamma_{k}-1+\sqrt{4\gamma_{k}^{2}-3}}{2}$, $\frac{2\gamma_{k}-1-\sqrt{4\gamma_{k}^{2}-3}}{2}$, $k=1,2,\ldots,m$.  Since $\gamma_{k}\geq1$, so for each $k=1,2,\ldots,m$, one has $\frac{1+2\gamma_{k}+\sqrt{4\gamma_{k}^{2}-3}}{2}\geq0$, $\frac{1+2\gamma_{k}-\sqrt{4\gamma_{k}^{2}-3}}{2}\geq0$, $\frac{2\gamma_{k}-1+\sqrt{4\gamma_{k}^{2}-3}}{2}\geq0$, while $\frac{2\gamma_{k}-1-\sqrt{4\gamma_{k}^{2}-3}}{2}\leq0$.
It follows that $\|\Gamma_{S}+0_A\oplus i\Omega_B\|_1=\sum_{k=1}^{m}(1+2\gamma_{k}+\sqrt{4\gamma_{k}^{2}-3})+2(n-m)$.

Therefore, ${\mathcal J}_1(\rho)=\frac{\sum_{k=1}^{m}(1+2\gamma_{k}+\sqrt{4\gamma_{k}^{2}-3})+2(n-m)}{\sum_{k=1}^{m}(4\gamma_{k})+2(n-m)}-1,$ ${\mathcal J}_2(\rho)=\sum_{k=1}^{m}(1+2\gamma_{k}+\sqrt{4\gamma_{k}^{2}-3})-\sum_{k=1}^{m}(4\gamma_{k})
=\sum_{k=1}^{m}(1-2\gamma_{k}+\sqrt{4\gamma_{k}^{2}-3})$, as desired. It is clear that ${\mathcal J}_1(\rho)={\mathcal J}_2(\rho)=0$ if and only if $\gamma_{k}=1$ for all $k=1,2,\ldots,m$. So the last assertion of the theorem is true.

The proof for the case $m>n$ is similar. \hfill$\Box$

{\bf Proof of Theorem 4.9.} It is clear that ${\rm Tr}(\Gamma_0)=2(a+b)$.  Since $(\Gamma_{0}+0_A\oplus i\Omega_B)^{\dagger}=\Gamma_{0}+0_A\oplus i\Omega_B$,
$\|\Gamma_{0}+0_A\oplus i\Omega_B\|_1=|\lambda_{1}|+|\lambda_{2}|+|\lambda_{3}|+|\lambda_{4}|$, where $\lambda_{i}$, $i=1,2,3,4$, are the eigenvalues of $\Gamma_{0}+0_A\oplus i\Omega_B$. A calculation gives that
$$\lambda_{1}=\frac{a+b+1}{2}+\frac{\sqrt{(b-a+1)^{2}+4c^{2}}}{2},$$
$$\lambda_{2}=\frac{a+b+1}{2}-\frac{\sqrt{(b-a+1)^{2}+4c^{2}}}{2},$$
$$\lambda_{3}=\frac{a+b-1}{2}+\frac{\sqrt{(a-b+1)^{2}+4c^{2}}}{2},$$ $$\lambda_{4}=\frac{a+b-1}{2}-\frac{\sqrt{(a-b+1)^{2}+4c^{2}}}{2}.$$
Since $ab-1\geq c^{2}$, we always have $\lambda_{1}, \lambda_{2}, \lambda_{3}\geq0$. If $a(b-1)-c^{2}\geq0$, then $\lambda_{4}\geq0$, and thus $\|\Gamma_{0}+0_A\oplus i\Omega_B\|_{1}=\lambda_{1}+\lambda_{2}+\lambda_{3}+\lambda_{4}=2(a+b)={\rm Tr}(\Gamma_0)$. In this case, one has ${\mathcal J}_1(\rho)={\mathcal J}_2(\rho)=0$, that is,  $\rho\in{\mathcal US}_{A\rightarrow B}^{G}$. If $a(b-1)-c^{2}<0$, then $\|\Gamma_{0}+0_A\oplus i\Omega_B\|_{1}=\lambda_{1}+\lambda_{2}+\lambda_{3}-\lambda_{4}=1+a+b+\sqrt{(a-b+1)^{2}+4c^{2}}$, which gives ${\mathcal J}_1(\rho)=\frac{1+a+b+\sqrt{(a-b+1)^{2}+4c^{2}}}{2(a+b)}-1$ and
${\mathcal J}_2(\rho)=  1+\sqrt{(a-b+1)^{2}+4c^{2}}-(a+b) .$

Therefore, we obtain that
$${\mathcal J}_1(\rho)={\rm max}\{0, \frac{1+a+b+\sqrt{(a-b+1)^{2}+4c^{2}}}{2(a+b)}-1\}$$
 and
$${\mathcal J}_2(\rho)={\rm max}\{0, 1+\sqrt{(a-b+1)^{2}+4c^{2}}-(a+b)\}.$$ \hfill$\Box$

\end{document}